\documentclass[twocollum]{aa}

\usepackage{graphicx}
\usepackage{natbib}
\usepackage{float}
\titlerunning{Gravitational acceleration mapping}
\authorrunning{Guang-Xing Li et al.}
\usepackage{amsmath}
\usepackage{amssymb}
\makeatletter

\begin{document}
\title{Gravitational acceleration and edge effects in molecular
clouds\thanks{Colorblindness-proof versions of the figures can be found in the
Appendix.}} \author{Guang-Xing Li \inst{1} \and  Andi Burkert \inst{1, 2} \and Tom
Megeath
\inst{3} \and Friedrich Wyrowski \inst{4}}
\institute{University Observatory Munich, Scheinerstrasse 1, D-81679 M\"unchen,
	Germany
	\and Max-Planck-Fellow, Max-Planck-Institute for Extraterrestrial
	Physics, Giessenbachstrasse 1, 85758 Garching, Germany
	 \and University of Toledo, Ritter Astrophysical
 Observatory, Department of Physics and Astronomy, Toledo OH 43606
 Germany \and Max-Planck Institut f\"ur Radioastronomie, Auf dem H\"ugel, 69,
 53121 Bonn}

\offprints{Guang-Xing Li, \email{gxli@mpifr-bonn.mpg.de, gxli@usm.lmu.de}}
\abstract{ Gravity plays important roles in  the evolution of molecular clouds.
We present an acceleration mapping method to estimate the acceleration induced
by gravitational interactions in molecular clouds based on observational data.
We find that the geometry of a region has a significant impact on the behavior
of gravity.
 In the Pipe nebula which can be approximated as a gas filament, we find that
 gravitational acceleration can effectively compress the end of this filament,
 which may have triggered star formation. We identify this as the ``gravitational focusing'' effect proposed by
Burkert \& Hartman (2004). In the sheet-like IC348-B3 region, gravity can lead
to collapse at its edge, while in the centrally condensed NGC1333
cluster-forming region gravity can drive accretion towards the center. In general, gravitational acceleration tends to be enhanced
in the localized regions around the ends of the filaments and the edges of
sheet-like structures. Neglecting magnetic fields,
these ``gravitational focusing'' and ``edge collapse'' effects can promote the
formation of dense gas in a timescale that is much shorter than the global dynamical time. Since the interstellar medium is in
general structured,  these edge
effects should be prevalent.
} 
\maketitle

\section{Introduction}

Star formation takes place in the dense and shielded parts of the molecular
interstellar medium (ISM) \citep{2014prpl.conf....3D}, and it is probably
determined by a combination of turbulence \citep{2004RvMP...76..125M,2012A&ARv..20...55H},
gravity \citep{2009ApJ...699.1092H,2011MNRAS.411...65B,2012MNRAS.427.2562B},
magnetic field \citep{2014prpl.conf..101L} and ionization radiation
\citep{1994A&A...290..421W,2009MNRAS.398.1537D}.

Gravity is one of the fundamental forces in nature, and it plays a
determining role in the formation of the stars. However, observational
understandings of its importance are still limited.
Various methods have been proposed to quantify the importance of
gravity on local \citep{1992ApJ...395..140B} and global \citep{2015A&A...578A..97L} as well
as various immediate scales \citep{2008ApJ...679.1338R,2009Natur.457...63G}.
In most of these methods, the effect gravity is estimated from the
\emph{bulk} properties of the regions obtained through averaging (e.g. the
total mass and bulk velocity dispersion). {  In contrast to these, we
propose a new method to study how gravity acts on individual structures.}

It is expected that detailed morphologies of the gas are tightly linked to
the star formation activities.
This can be illustrated by the simulations of
\citet{2004ApJ...616..288B,2007ApJ...654..988H}. In a simplified case, the
collapse of a roundish disk of a constant surface density under the influence of self-gravity is simulated.
Different from the common perception that the disk
will merely shrink in radius, the simulation shows that matter accumulates at
its edge. Such an accumulation of gas leads to subsequent formation of filaments
at the edge and finally dense cores. The collapse
of filaments are also simulated \citep[see also][]{2015MNRAS.449.1819C,
2015MNRAS.452.2410S}, and it is found that matter accumulates at the ends of the
filaments.
To understand these phenomena, one needs to understand how gravity acts on gas
in regions with different geometries. These information can not be obtained
through e.g. the virial parameter. A better method which can provide estimates
 on the importance of gravity in regions with arbitrary geometries is thus
 needed.

There have been growing interests in studying filamentary
structures in molecular clouds \citep{2010A&A...518L.102A, 2011A&A...529L...6A,2012A&A...541A..63P,2014A&A...561A..83P,
 2013A&A...550A..38P, 2010A&A...518L.106K, 2012A&A...540L..11S,
 2014prpl.conf...27A}. Both numerical and analytical approaches have been employed to understand the evolution of these structures
 \citep{2012ApJ...744..190T, 1995ApJ...438..226T} \citep{2014MNRAS.445.2900S,
 2015MNRAS.452.2410S, 2015MNRAS.449.1819C} \citep{2014MNRAS.443..230H,
 2015MNRAS.446.2110T, 1992ApJ...388..392I, 2015arXiv150400647W}.
 Filamentary structures have also been identified at $ 10^2 - 10^3 \;\rm
 pc$ scale \citep{2013A&A...559A..34L, 2014A&A...568A..73R,
 2014MNRAS.441.1628S, 2014ApJ...797...53G,2015arXiv150400647W}. Due to their highly irregular
 morphologies, we expect gravity to behave in a non-uniform way in these
 regions.

In this paper, we introduce an acceleration mapping method to quantify the
effect of gravity on molecular ISM.
Our general aim is to provide intuitive and qualitative pictures of the strength
of gravitational acceleration in molecular clouds. Section
\ref{sec:method} we introduce the method. The effect of gravity in accelerating
gas is discussed in Sec. \ref{sec:motion}. Then we apply it to idealized
examples as well as real observational data (Sections \ref{sec:appl}, \ref{sec:obervation}).
The importance of gravitational acceleration in star formation is discussed in
detail in Sec.  \ref{sec:disc}.
In Section \ref{sec:conclusec} we conclude.

\section{Calculating acceleration}\label{sec:method}
\subsection{General concept}\label{sec:concept}
Observationally, gas is observed on the sky plane, and the distribution of gas
along the line of sight is not easily observed. To compute gravitational
potential, we assume that all the
observed matter is distributed in a thin plate of thickness $H$.
 We make this assumption mainly because of observational limit, namely that matter can only be
reliably mapped on the sky plane in most observations.
A generalization of the method to the 3D case is straightforward.

\begin{figure}[H]
\includegraphics[width =  0.45 \textwidth]{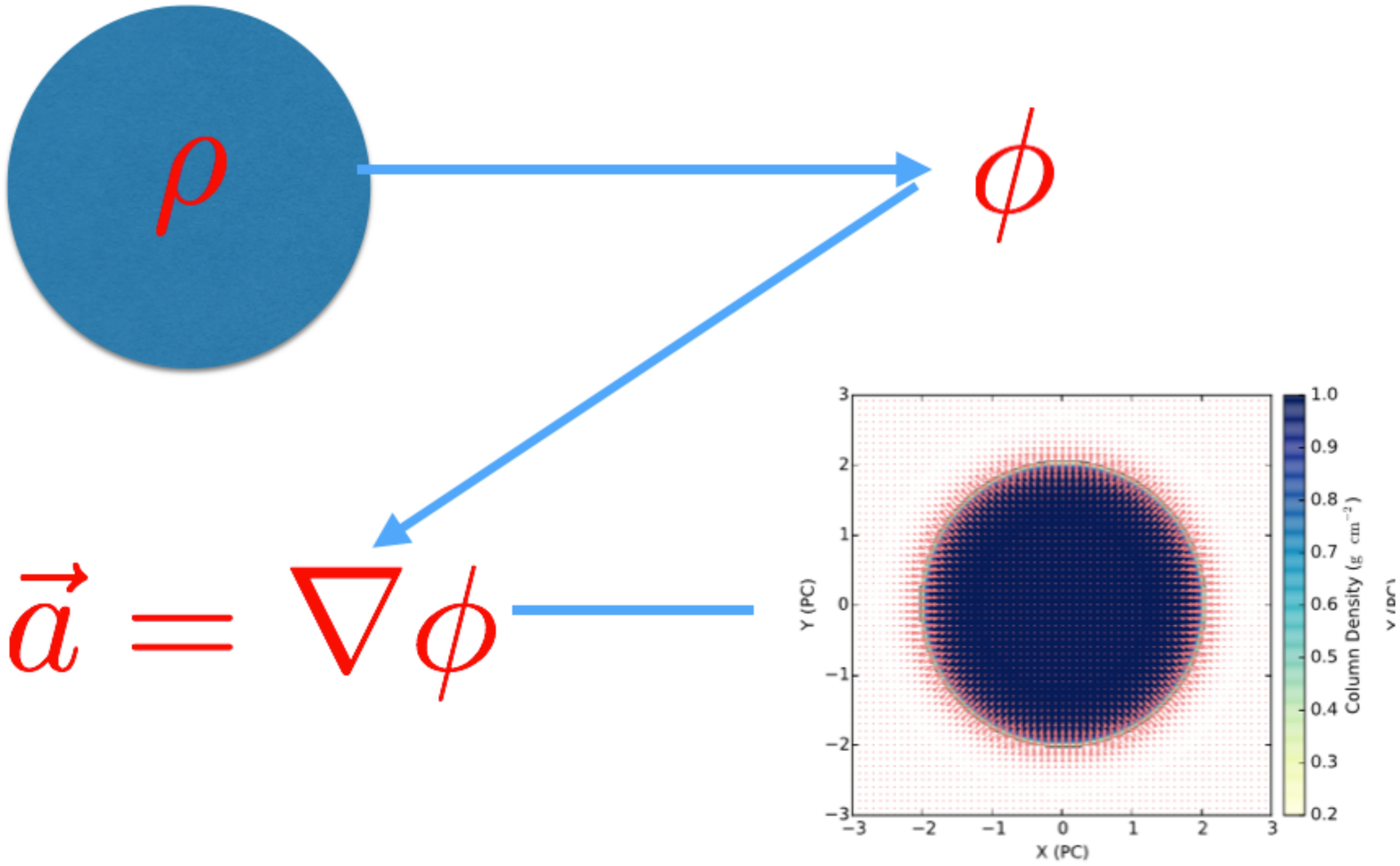}
\caption{A illustration of our concept. See Sec. \ref{sec:concept} for
details. { $\rho$ stands for density, $\phi$ stands for gravitational
potential, and $\vec{a}$ stands for acceleration.}
\label{fig:prededure}}
\end{figure}

In our work, we first compute gravitational potential based on a column
density distribution. Then we compute gravitational acceleration distribution
based on the gravitational potential. Finally, we combined the gravitational
acceleration map with the column  density map, and study the effect of gravity on matter.
This is illustrated in Fig.
\ref{fig:prededure}.

\subsection{Numerical realization}
Observationally, matter can be observed on the sky plane as $\Sigma(x, y)$
where $\Sigma$ is the surface density. In our study, we compute
gravitational acceleration based on $\Sigma(x, y)$.

Assuming the observed gas is distributed on the 2D sky plane with a thickness
 $H$, the projected gravitational potential can be computed as
\citep{2011ApJ...729..120G}
\begin{equation}\label{eq:gong}
\Phi_{\vec{k}, \rm 2D} = \frac{- 2 \pi \; G \; \Sigma_{\vec{k,  \rm
2D}}}{|\vec{k}| (1 + |\vec{k} H|)}\;,
\end{equation}
where $\Phi_{\vec{k}, \rm 2D}$ is the projected gravitational potential in the
$\vec{k}$ space, and $\Sigma_{\vec{k}, \rm 2D}$ is the surface density in the
$\vec{k}$ space.
Numerically, we first make a Fourier
transform to $\Sigma(x, y)$ and obtain $\Sigma_{\vec{k}}$, then use Eq.
\ref{eq:gong} to evaluate $\Phi_{\vec{k}, \rm 2D}$. Finally, $\Phi(x, y)$ is
obtained through an inverse Fourier transform of $\Phi_{\vec{k}, \rm 2D}$.

 In reality,
gravitational potential is a quantity defined in the 3D space.
In this formalism, at one position on the sky plane $(x_0, y_0)$, the
2D projected gravitational potential $\Phi(x_0, y_0)$  is defined as $\Phi(x_0,
y_0) = \Phi(x_0, y_0, z)$ where $z$ represent the direction of the line of sight
and $z = 0$ .

With the gravitational potential, acceleration can be derived:
\begin{equation}
\vec{a} =  - \nabla \Phi \; .
\end{equation}

The gradient is evaluated with \texttt{np.gradient} from the
$\texttt{numpy}$ package.

{  The only parameter we need to introduce is the thickness along the $z$
direction $H$. In general, when $H$ is small, the structure of smaller scales
are better represented, and when $H$ is large, small-scale structures tend to be
smoothed out in the acceleration map. Ideally, $H$ should be chosen to be
close to the real 3D thickness of the cloud. However, this is not feasible in
practise. Here, we choose to use $H =0.3 \;\rm pc$ throughout this paper. This is because
observations have demonstrated that star formation tend to concentrate in
clumps, whose sizes are close to parsec scale
\citep{2002ApJ...566..945B,2000A&A...355..617M}. The size of the clumps provides
an hint to the expected thickness of the cloud. On the other hand, we do not choose a $H$
that is much larger than a parsec, because we would still like to study the
small-scale structure of gravitational acceleration.
}

{
\subsection{Boundary conditions}\label{sec:padding}
Computing gravitational potential and acceleration in the Fourier space
automatically assumes periodic boundary condition. This means it is assumed that
the input density distribution $\Sigma(x, y)$ repeats itself in the $x$--$y$
plane ( $\Sigma(x, y) = \Sigma(x + x_0, y) = \Sigma(x, y + y_0) = \Sigma(x +
x_0, y + x_0)$ when the image have a size of (x0, y0)).

Since gravity is a long-range force, the presence of matter outside
the box do have influences on the motion of gas inside the box. Therefore, the
results are dependent on the boundary conditions which have to be specified.

In this work, for an input map of size $(x_0, y_0)$ we added zeros to the
region where $x < 0$, $x > x_0$, $y < 0$ or $y > y_0$, and created a map of the
size  $(2 \times x_0, 2\times y_0)$. Then we use Eq. \ref{eq:gong} to compute
the gravitational potential. This is equivalent to assuming that the cloud is
surrounded by a region devoid of matter. The size of the void region is
comparable to the cloud size. One can also increase the size of this void
region. Because gravitational force decays as $r^{-2}$, our results are not
sensitive to this. However this will increase the computational cost
significantly.

Nature is composed of a continuous distribution of matter, which are
gravitationally interacting (e.g. molecular clouds are in constant gravitational
interaction with the spiral arms).
In computing our acceleration map, one has to isolate the object of interest and
study the gravitational acceleration arising from the gravity originates for this
object. This isolation procedure can also lead to inaccuracies. However, one can
always evaluate these effects by comparing the expected gravitational
acceleration from a body that is outside our box ($|\vec{a}| \approx G m /r^2$ where $m$ is
the mass of the object outside our box, and $r$ its the distance from our object) with the acceleration computed
in our map, to evaluate the effect of the simplified boundary conditions on our
results.

}

\subsection{Projection effects}
The acceleration map we obtain with this method is computed at $z = 0$. In
reality matter are not distributed in a plane but are distributed in 3D. This
will have an effect on our results. In general, acceleration maps are affected
by projection effects, and cautions should always be taken when the region has a
complicated structure or a multiple of velocity components. {  Here we briefly
discuss various projection effects. }

 {
 Because of line-of-sight contamination, physically unassociated
 structures can appear as coherent. This will certainly have an impact on our
 results.
  However, in many
cases, the line-of-sight contamination can be accessed using 3D extinction
\citep{2014MNRAS.443.2907S}
\citep{2015arXiv150701005G,2014MNRAS.443.1192C,2014Sci...344..183K}
\citep{2014MNRAS.438.2938H,2011AAS...21725008B,2014A&A...561A..91L} or velocity
information.
 }

{
For clumpy structures, the fact that we can not distinguish structures along the
line of sight only have a moderate influence on the results.
 Consider a portion of a clump that is separated from the center by a distance
 $L$.
 In our calculations, it is automatically assumed that the all the gas stays at
 $z=0$.
 If a portion of the cloud with mass $m$ stays at $z = z_0 $ instead of $z = 0$, the acceleration it
 contributes is $a \sim G m / (L^2 + z_0^2)$. For a clump, because the geometry
 is close to symmetric, $z_0 \lesssim L$, the error is relatively small, and is
 generally acceptable.

For filamentary structures, inclination has a significant impact on the result.
To illustrate this one can look at the gravitational acceleration around a
filament. Consider a filament of length $l_0$ and width $d$. The filament has a
 line mass of $\delta = {\rm d} m /
{\rm d} l$. The gravitational acceleration at one end of this filament can be
solved approximately:
\begin{equation}\label{eq:filament}
a = \int_{d}^{l_0} \frac{G \delta }{r^2}{\rm d} r = G \;\delta\;
\Big{(}\frac{1}{d} - \frac{1}{l_0}\Big{)}\;,
\end{equation}
where $G$ is the gravitational constant.
Consider a case where the filament is not parallel to the sky plane but is
inclined by an angle
$\theta$. From the observer, $\delta' = \delta / \cos(\theta)$, $d' = d
\,\cos(\theta) $ and $l' = l_0 \,\cos(\theta)$. Overall, according to Eq.
\ref{eq:filament}, $a' = a\, /\cos^2(\theta)$. If the filament is inclined by
$45^{\circ}$, the gravitational acceleration will be over-estimated by a factor
of $2$. A larger inclination leads to a larger error.

For sheet-like structures, inclination has some impacts on the results but these
are in general acceptable. This can be illustrated by some examples in
\citet{2004ApJ...616..288B} where the ellipse case shows a good similarity
with the case of the roundish disk. Since we only expect our results to be
accurate in the order-of-magnitude sense, the error arising from the
inclination effect is in general acceptable.

To sum up, all the projection effects leads to some over-estimations
of the acceleration.
The line-of-sight contamination can be avoided in many cases using velocity
information or extinction distance estimates. For clump-like structures,
line-of-sight effects only
moderately affect the computed acceleration structure.
For filaments and sheets, a moderate inclination can often lead to a change in
the estimated acceleration by a factor of 2, which is acceptable. The acceleration
map computed in this work should thus capture the general behaviors of
gravity in molecular clouds, but the values in the acceleration
map is accurate only in the order-of-magnitude sense.

}

{
\section{Acceleration and gas motion}\label{sec:motion}
Gravitation acceleration is important for the movement of gas. Consider that at
$p_1$ the acceleration is $\vec{a}_1$ and at $p_2$ the acceleration is
$\vec{a}_2$. Initially, both the gas at position $p_1$ an the gas at position
$p_2$ have zero velocities. We are interested in the effect of acceleration in gas compression / dilation.

At time $t$, the different in separation is simply
\begin{equation}
\delta x = \frac{1}{2}|\vec{a_2} - \vec{a_1}| \times t^2\;.
\end{equation}
What actually drives the gas compression, dilatation or shear is the (vector)
difference in acceleration.

In a region, when the acceleration is larger,
it is very likely that there exist significant contrasts in
acceleration with respect to its surroundings. In this work, we choose to
study the acceleration rather than the difference in acceleration for two
reasons: First, in many cases, they are related. A large acceleration is
usually related to a large acceleration difference.
Second, there are other forces such as magnetic forces that are influencing the
cloud evolution.
To properly evaluate the impact of gravitational acceleration one need to
compare the acceleration strength with e.g. pressure from the magnetic field.
For this purpose, it is more practical to study acceleration instead of
the difference in acceleration.
}

\section{Applications to idealized examples}\label{sec:appl}
We apply the acceleration mapping
method to several idealized examples. We consider acceleration from a thin
disk, a filament and a centrally condensed gas clump.

\subsection{Thin disk}\label{sec:disk}
\begin{figure*}
\includegraphics[width = 0.95 \textwidth]{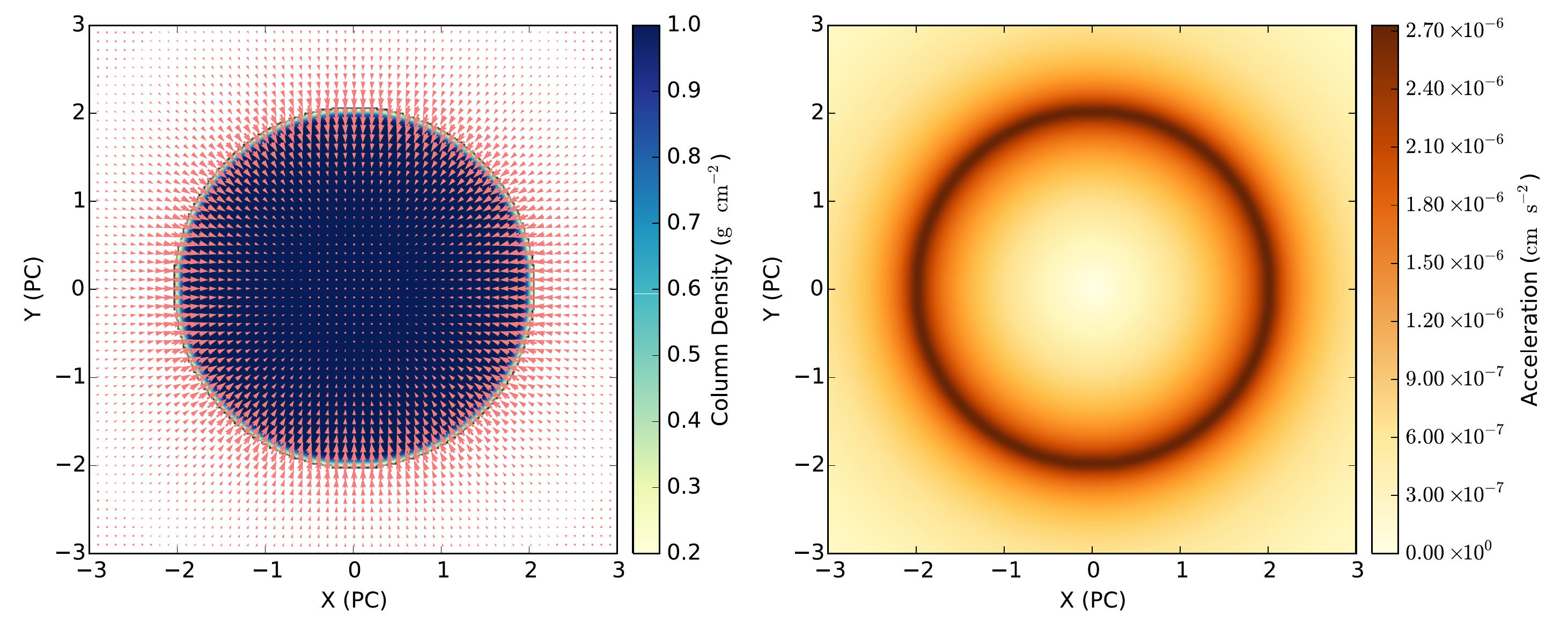}
\caption{{  Left Panel:} Density distribution and acceleration of our disk
model with a radius of $1\;\rm pc$ and a
surface density of $1\;\rm g \;\rm\; cm^{-2}$. The background image is the
density distribution and the vectors stand for acceleration. {  Right
Panel:} A map of the magnitude of acceleration. A color-blindness-proof version of the this figure and be found in
Appendix \ref{sec:appendix:a}.
\label{fig:disk} }
\end{figure*}

In our thin disk example, we consider a disk of a radius of $1\;\rm pc$ and a
surface density of $1\;\rm g \;\rm\; cm^{-2}$. The input is smoothed with a
Gaussian kernel with $\sigma = 2 \rm\; pixels$, and we use $H = 0.3\; \rm pc$
(see Eq. \ref{eq:gong}) throughout this paper. The results are presented in Fig.
\ref{fig:disk}.
In the disk case, the acceleration reaches a maximum at the edge of the disk.
Under ideal situations, matter should accumulate at the regions where
acceleration is enhanced, which was observed in \citet{2004ApJ...616..288B}.

\subsection{Truncated filament}\label{sec:filament}
\begin{figure*}
\includegraphics[width = 0.95 \textwidth]{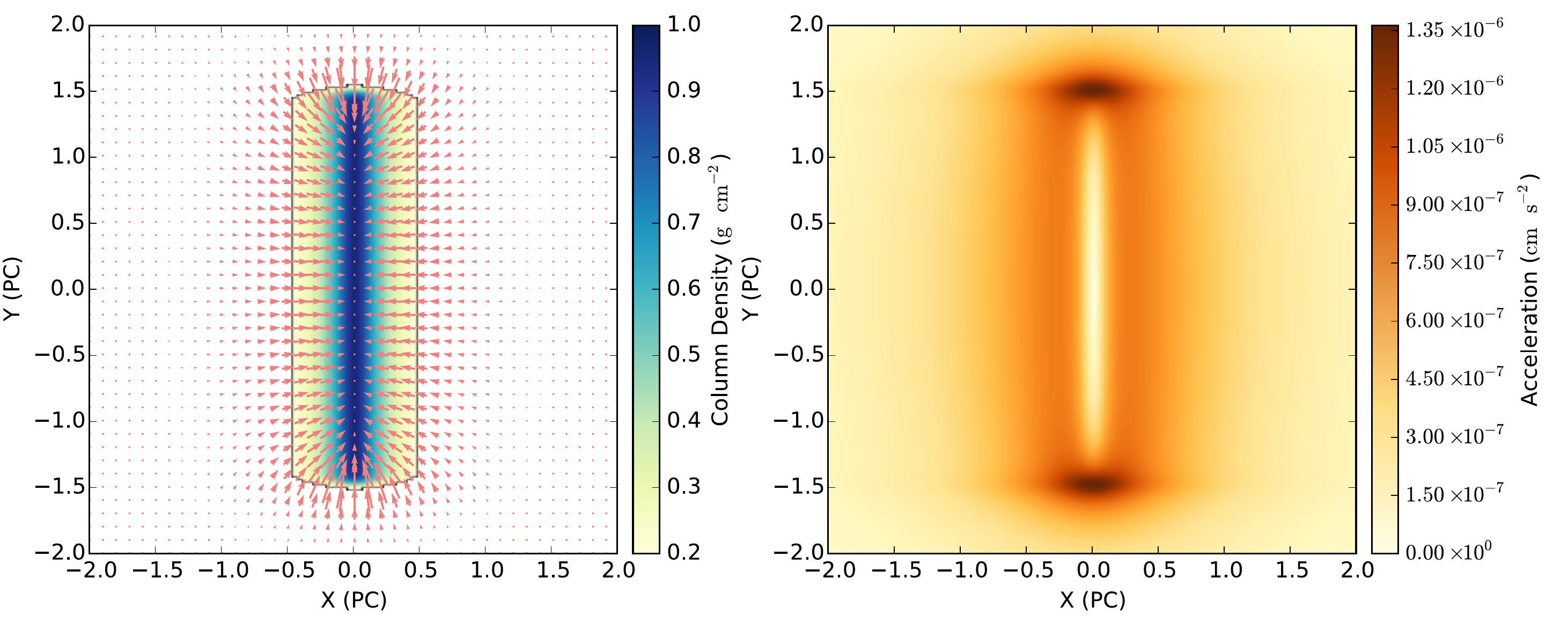}
\caption{{  Left Panel:} Density distribution and acceleration of our
truncated filament model (see Section \ref{sec:filament} for details). The
background image is the density distribution and the vectors stand for acceleration.
{  Right Panel:} A map of the magnitude of acceleration. A color-blindness-proof version of the this figure and be found in
Appendix \ref{sec:appendix:a}.
\label{fig:filament} }
\end{figure*}

We consider a filament with a width of $0.1\;\rm pc$, a length of
$1\;\rm pc$, and a peak surface density of $1\,\rm g\rm\; cm^{-2}$.
{  Along the filament, the peak surface density is constant. Perpendicular to
the filament, we assume a Plummer-like density
profile}:
\begin{equation}\label{eq:filament}
\Sigma(r) = \Sigma_{\rm c}\frac{1}{1 + (l/l_{\rm flat})^{{p}/2}}\;,
\end{equation}
{  where $l$ is the projected distance to the ridge of the filament.
 We take
$p=1.6$, $l_{\rm flat} = 0.033\;\rm pc$ and $\Sigma_{\rm c} = 1 \rm\; g \;
cm^{-2} $. Integrating over Eq. \ref{eq:filament} gives a line mass  of 1.27
$\;\rm g\,cm^{-2}\times pc$.} The input is smoothed with a Gaussian kernel with $\sigma
= 2 \rm\; pixels$. $H = 0.3 \rm\; pc$. The results have been shown in Fig. \ref{fig:filament}. The acceleration reaches its maximum at both ends of the filaments, and this should leads to collapse.
This effect was observed in \citet{2004ApJ...616..288B} and
was named ``gravitational focusing'' by the
authors. {  Using Eq. \ref{eq:filament} and set $d = l_{\rm flat} = 0.033
\;\rm pc$, $l_0 = 1\;\rm pc$ and $\delta = 1.27\;\rm g\,cm^{-2}\times pc$ gives $a \approx 2 \times 10^{-6}\;\rm
cm\, s^{-2}$, in qualitative agreement with the results in Fig. \ref{fig:filament}.

}

Gravity is also responsible for driving accretion onto dense structures.
\citet{2013ApJ...769..115H} investigated the accretion onto dense
filaments, and identified fan-like structures of gas velocities on filament
ends.
Such fan-like velocity structures resemble acceleration vectors in Fig.
\ref{fig:filament}.

\subsection{Centrally condensed clump}\label{sec:clump}
\begin{figure*}
\includegraphics[width = 0.95 \textwidth]{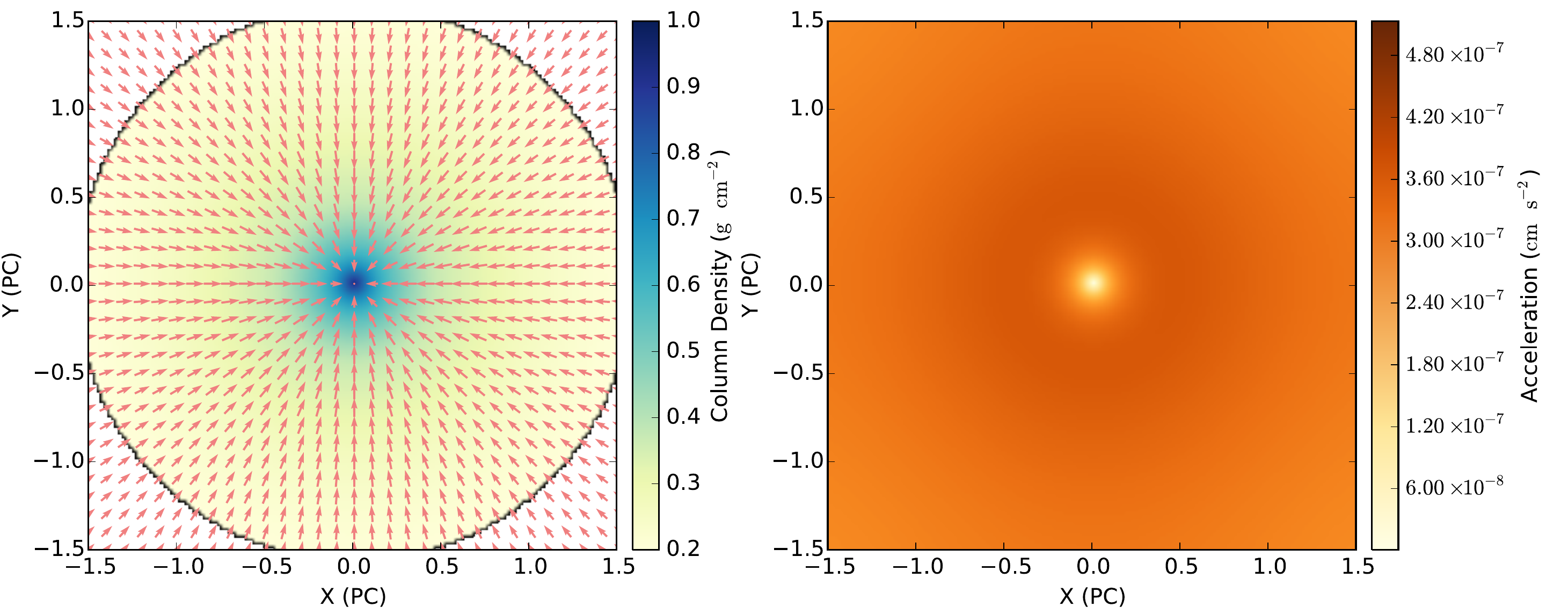}
\caption{{  Left Panel:} Density distribution and acceleration of our clump
model (see Section \ref{sec:clump} for details). The background image
is the density distribution and the vectors stand for acceleration. {  Right
Panel:} A map of the magnitude of acceleration. A color-blindness-proof version of the this figure and be found in
Appendix \ref{sec:appendix:a}.
\label{fig:clump} }
\end{figure*}

We consider the effect of gravity on a centrally condensed structure which
resembles those seen in observations
\citep[e.g.][]{2010ApJ...716..433K,1981MNRAS.194..809L}.
We assume
\begin{equation}
\frac{\Sigma}{M_{\odot} \; \rm pc^2} = 870 \times \Big(\frac{r}{\rm pc}+ 0.1
\Big)^{-0.7}\;,
\end{equation}
where $\Sigma$ is the surface density and $r$ is the projected distance from the
center. It mimic the structure of a gas condensation seen in
observations.
The gravitational acceleration structures are shown in Fig. \ref{fig:clump}. Here
acceleration vectors converge to the center where the surface density reaches its
maximum, and in this case, it will probably lead to a continuous accretion flow.
The magnitudes of acceleration vectors are very small at the central region.
This is because of symmetry, which leads the acceleration from the outer
parts of the clump to cancels out.

\section{Applications to observational data}\label{sec:obervation}
After testing the method with the idealized cases, we apply the method to
observational data.

\subsection{ Pipe Nebula}\label{sec:pipe}
We obtain extinction maps from
\citet{2009MNRAS.395.1640R,2010MNRAS.406.1350F}. We use the 25 stars $A_{\rm v}$
maps which have a pixel size of 0.5 $\rm arcsec$. The extinction $A_{\rm v}$
and surface density are linked by \citep{1978ApJ...224..132B}
\begin{equation}
\frac{n({\rm H_2})}{A_{\rm v}} = 9.4 \rm \times {10^{20}\;cm^{-2}\; mag^{-1}}\;.
\end{equation}

The Pipe nebula locates at a
distance of $\sim 160\;\rm pc$ \citep{2007A&A...470..597A,
1998A&A...338..897K}. The cloud contains $\sim 10^4\;M_{\odot}$, and has an
elongated morphology. It is believed that such a structure is shaped by the
expansion of another H{\sc ii} region \citep{2012ApJ...754L..13G}. Only
$\sim 150\rm\; YSOs$ are found to be associated with the whole cloud
\citep{2009A&A...508L..35K,2006A&A...454..781L}.
 Filamentary
structures in the nebula have been found in recent \texttt{Herschel} observations \citep{2012A&A...541A..63P}.

Star formation in the Pipe nebula is relatively quiescent, except the B59
region where a cluster of YSOs can be found
\citep{2007ApJ...655..364B,2010ApJ...719..691F, 2009ApJ...704..292F}. The
physical reason for this uneven distribution of star formation is still
unclear. It is proposed that the star formation
activity is triggered by compression  from the nearby Sco OB2
association \citep{1999PASJ...51..871O,2012A&A...541A..63P}. Magnetic support
might also play a role \citep{2008A&A...486L..13A}.

The Pipe nebula is an ideal target for our analysis, for two reasons. First, the
region have a relatively high galactic latitude, which makes it relatively easy
to separate it from the background. Second, the region has an elongated
morphology, making it an ideal testbed for non-local effects of gravity.

\begin{figure*}
\includegraphics[width =  0.95 \textwidth]{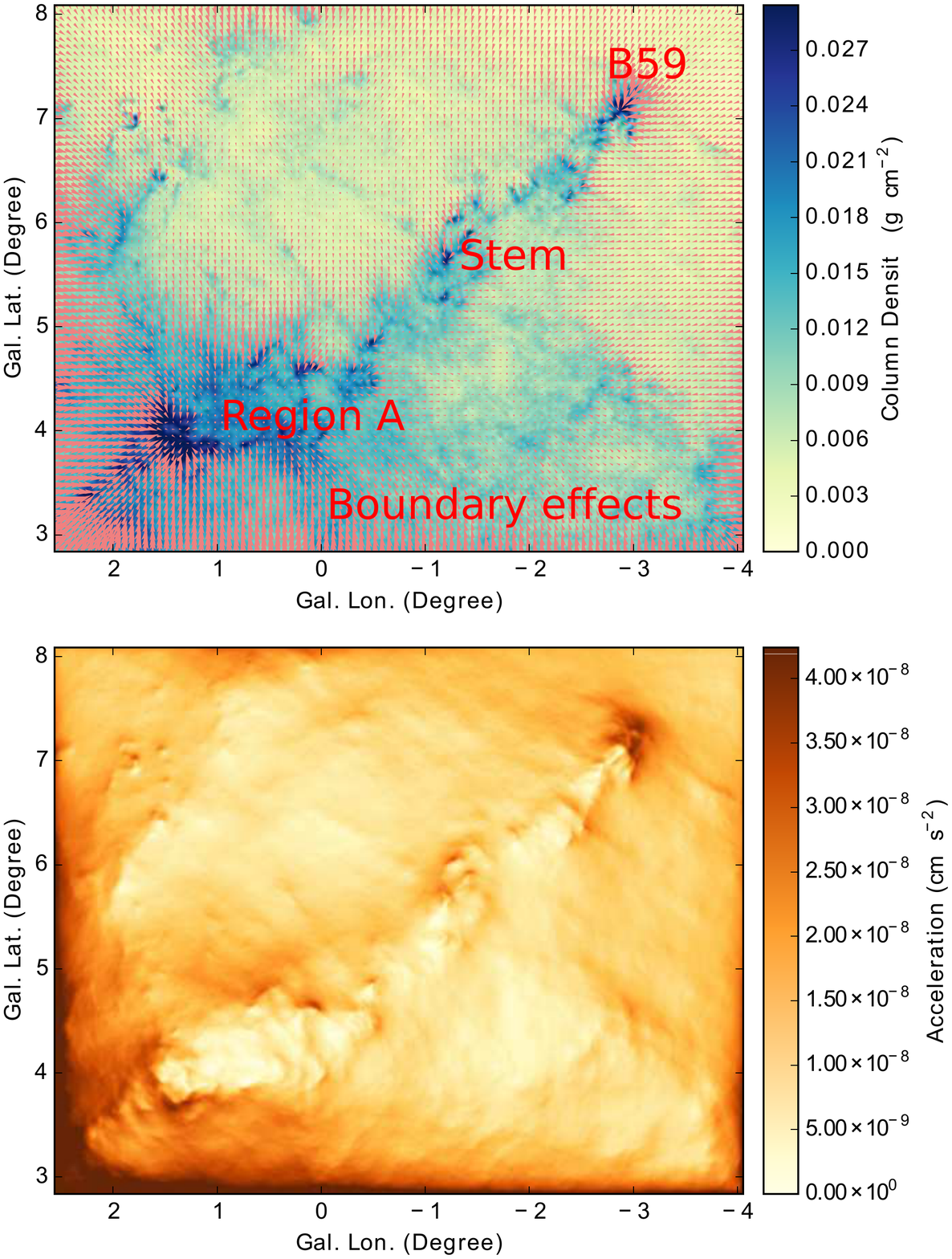}
\caption{{  Upper Panel:} surface density distribution and acceleration of the
Pipe region. The background image is the surface density distribution and the
vectors stand for acceleration. {  Lower Panel:} A map of the magnitude of
acceleration of the same region. The region A, ``stem'' region and B59 regions
are indicated. Regions where the artifacts due to boundary effect are also
indicated. Here one degree correspond to $\sim 3\;\rm pc$.
A color-blindness-proof version of the this figure and be found in
Appendix \ref{sec:appendix:a}.
\label{fig:pipe:region} }
\end{figure*}

In Fig. \ref{fig:pipe:region} we present surface density and acceleration
structure of Pipe nebula. The extra acceleration on the lower left edge of the
map is due to artifacts, since the nebula have a finite surface density inside
the map, and outside the map it is assumed that the surface density is zero. {  Because
 sharp edges are introduced around the boundaries, artificial accelerations
arise from the computations (see Sec. \ref{sec:padding})}.  Therefore
these boundary regions are excluded from our analysis.
The acceleration has been enhanced at region A (see Fig. \ref{fig:pipe:region}) and the B59 region,
and together with the ``stem'' region, these three form a system where
region A and the B59 region are the ends of a truncated filament and the
``stem'' region correspond to the ``body'' of a filament, and accelerations are
enhanced at the ends of this filament. The enhancement of acceleration at the ends results from
the gravitational focusing effect observed in \citet{2004ApJ...616..288B} and
 in Section \ref{sec:filament}.

\begin{figure}
\includegraphics[width = 0.5 \textwidth]{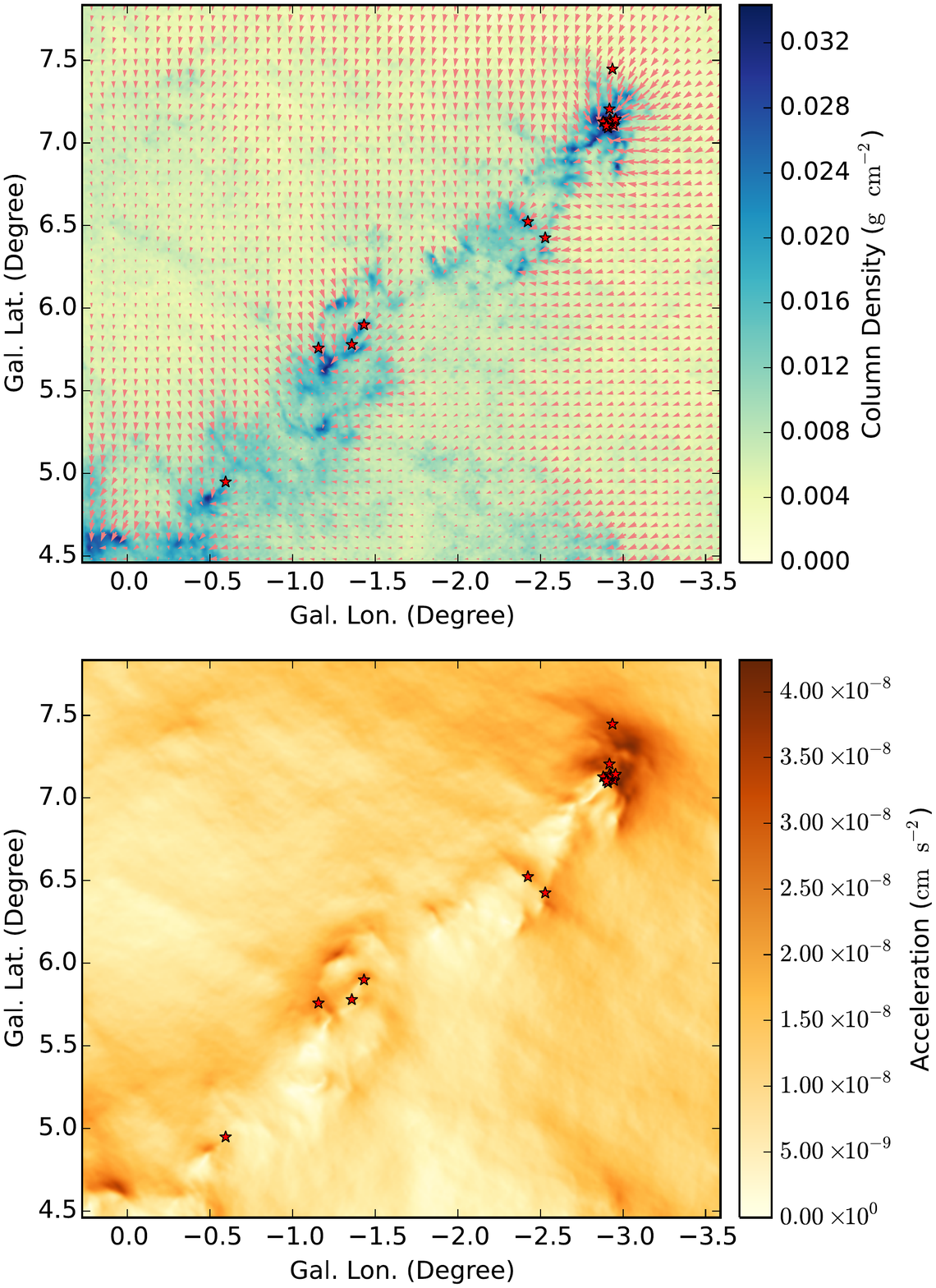}
\caption{{  Upper Panel:} surface density distribution and acceleration of the
Pipe Nebula. The background image is the surface density distribution and the
vectors stand for acceleration.
{  Lower Panel:} A map of the magnitude of acceleration of the same region.
The red stars stand for the YSO candidates in \citep{2009ApJ...704..292F}.  Here one degree correspond to $\sim
3\;\rm pc$. A color-blindness-proof
version of the this figure and be found in Appendix \ref{sec:appendix:a}.
\label{fig:pipe1} }
\end{figure}

In Fig. \ref{fig:pipe1} we present surface density and acceleration structure
of the stem and B59 region of the Pipe Nebula. YSO candidates from
\citet{2009ApJ...704..292F} are overlaid.
The B59 region shows up as a region where acceleration vector
 converges; this is also the region where the majority of the protostars are
 found.  This implies a connection between gravitational acceleration and
 star formation.

\begin{figure*}
\includegraphics[width = 0.9 \textwidth]{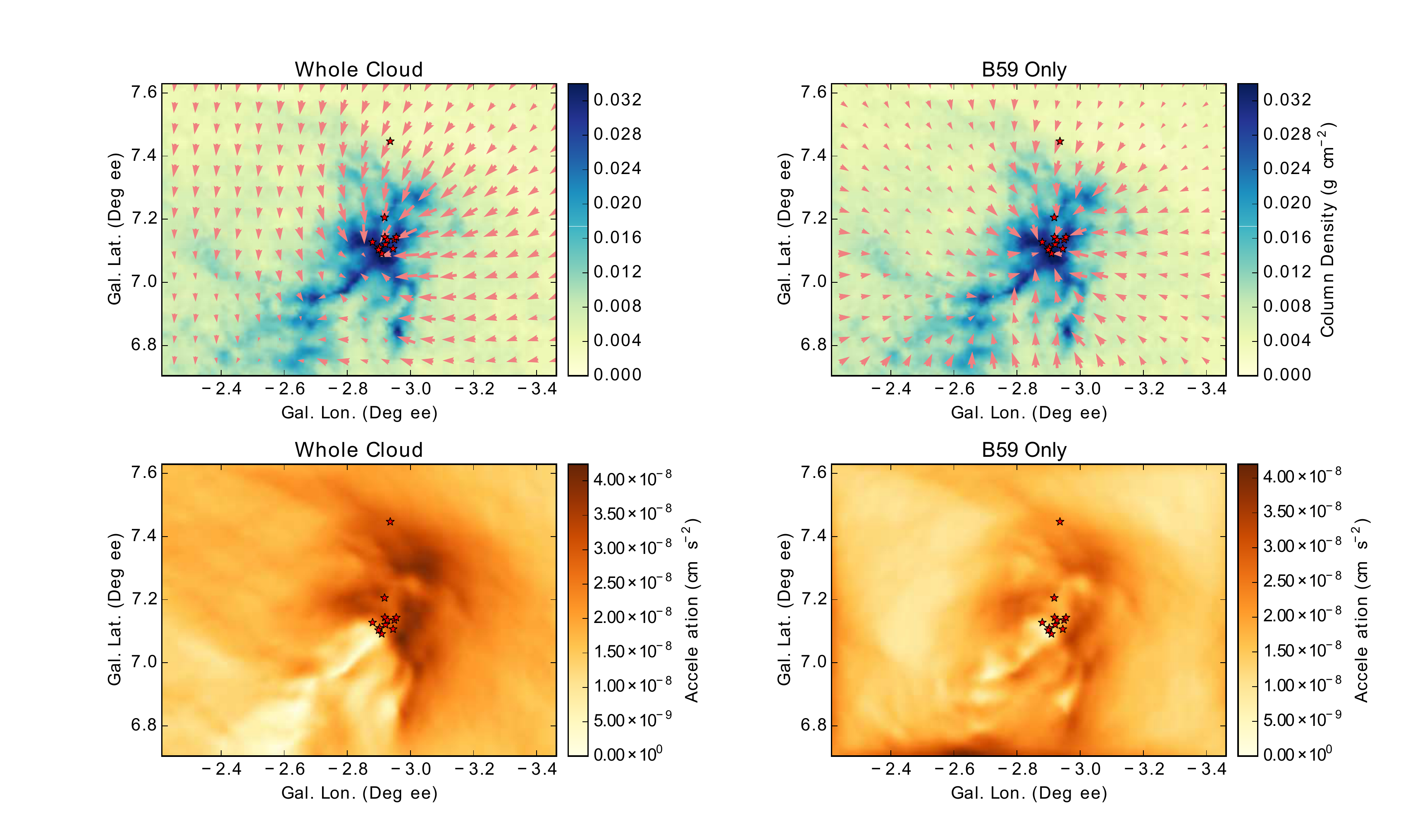}
\caption{{  Upper Panels:} surface density maps and acceleration of the
B59 region in the Pipe Nebula. The background images are the surface density
distribution and the vectors stand for acceleration.
{  Lower Panels:} Maps of the magnitude of acceleration of the same region.
The red stars stand for the YSO candidates in \citet{2009ApJ...704..292F}.
The results shown in the left panels are computed by taking all the
matter of the Pipe nebula into account \emph{(whole cloud)}, and the
results shown in the right panels are computed by taking only the matter
inside the B59 into account \emph{(B59 only)}.  Here one degree correspond to
$\sim 3\;\rm pc$. \label{fig:pipe2} A color-blindness-proof
version of the this figure and be found in Appendix \ref{sec:appendix:a}.}
\end{figure*}

 The acceleration seen around the B59 region have two possible sources. The
 first and obvious source of acceleration is the gravity from the central part
 of the B59 where the surface density peaks. The second source of acceleration is the
 gravity contributed from the rest of the Pipe nebula.
 To further investigate the origin of acceleration we  \emph{repeat the
 calculation by considering only the gas inside the B59 region, and compare the
 results with acceleration map computed for the whole Pipe nebula. }The
 results are presented in Fig. \ref{fig:pipe2}.
 Several clear differences can be identified.
 First, acceleration computed for the B59 region alone converges to
 its center, and when the gravity of the whole Pipe nebula has been taken into
 account, acceleration only converges through the north-western part, and
 acceleration at the south-eastern part of the B59 region is suppressed.
 This difference is due to the contribution from gas the belongs to the
 Pipe nebula but outside the B59 region. Second, the magnitude of acceleration at the north-eastern part of the
 the B59 region is much higher when the gravity of the whole Pipe nebula have
 been taken into account, which indicates that gas from the whole Pipe
 nebula is contributing significantly to the gravitational acceleration observed in the
 B59 region. The
 large-scale acceleration can compress B59 region significantly.

 \subsection{ NGC1333 cluster-forming
 region}\label{sec:ngc1333} Centrally condensed gas clumps are common in
 star-forming regions. On such example is the cluster-forming region NGC1333 in the Perseus molecular cloud.

 We obtain extinction map of the Perseus molecular cloud from the COMPLETE
 \citep{2006AJ....131.2921R} survey.
 First, we compute the gravitational acceleration for the whole region. Then
 we zoom into the NGC1333 region and study the structure. A distance of 250 pc
 is used in our calculations \citep{1998A&A...338..897K}.

\begin{figure*}
\includegraphics[width = 0.95 \textwidth]{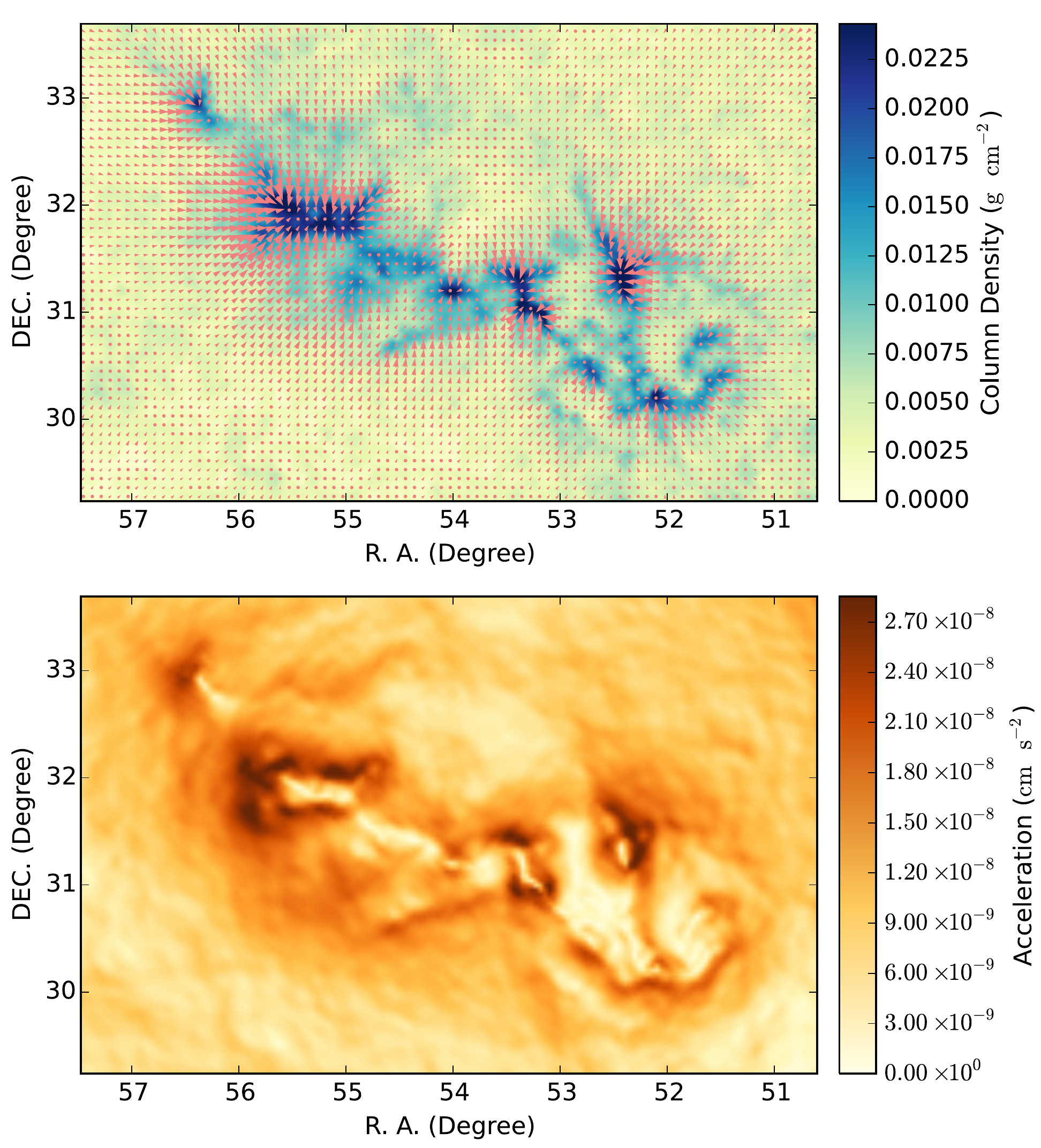}
\caption{{  Upper Panel:} surface density distribution and acceleration of the
Perseus molecular cloud. The background image is the surface density distribution and the
vectors stand for acceleration.
{  Lower Panel:} A map of the magnitude of acceleration of the same region.
Here one degree correspond to $\sim 4.5\;\rm pc$.
\label{fig:perseus_all} A color-blindness-proof
version of the this figure and be found in Appendix \ref{sec:appendix:a}.}
\end{figure*}

 The results for the whole Perseus region is shown in Fig.
 \ref{fig:perseus_all}, and the acceleration map exhibits a variety of
 behaviors in different regions.

\begin{figure*}
\includegraphics[width = 0.95 \textwidth]{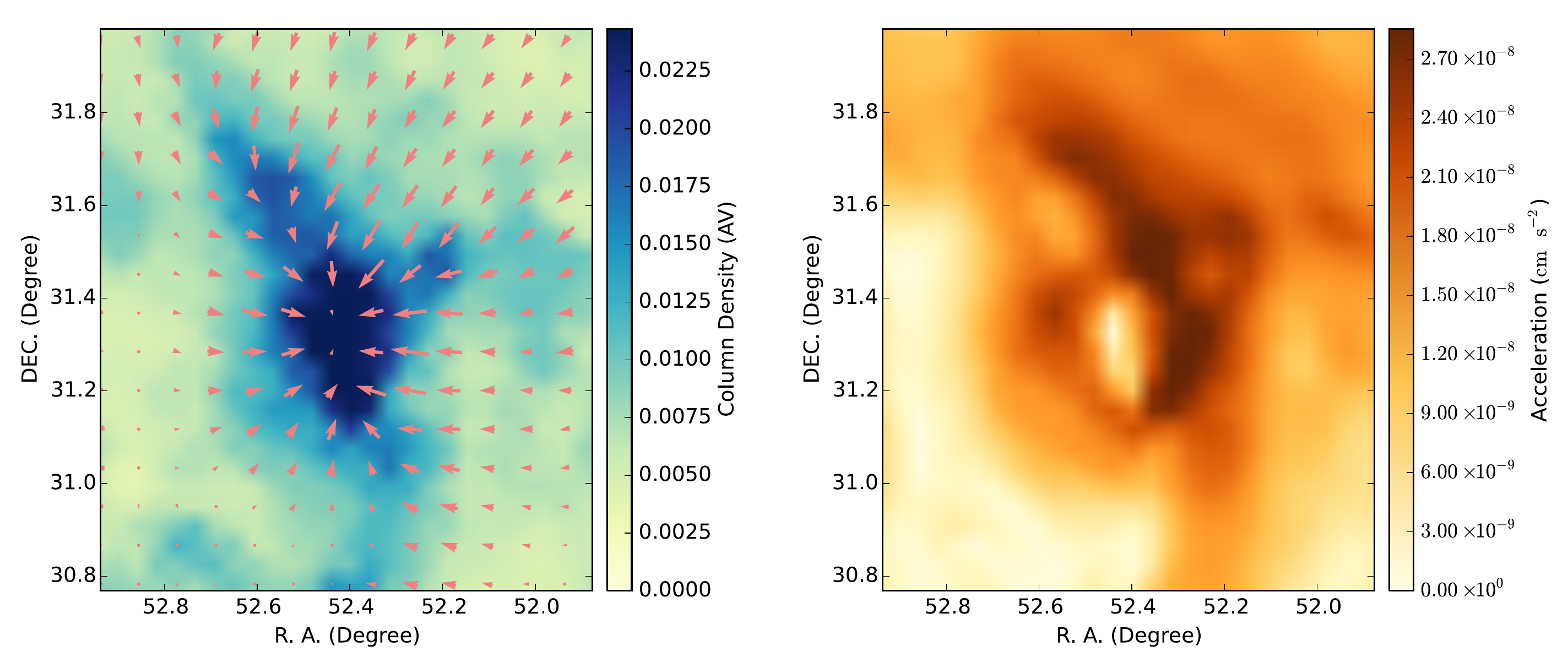}
\caption{{  Left Panel:} surface density distribution and acceleration of the
NGC1333 region in the Perseus molecular cloud. The background image is the surface density distribution and the
vectors stand for acceleration.
{  Right Panel:} A map of the magnitude of acceleration of the same
region. Here one degree correspond to $\sim 4.5\;\rm pc$.
\label{fig:ngc1333}A color-blindness-proof
version of the this figure and be found in the Appendix \ref{sec:appendix:a}. }
\end{figure*}

 The results for NGC1333 are presented in Fig. \ref{fig:ngc1333}. Here the
 acceleration has a relatively symmetric structure, and converges towards the
 center.
 This is mainly due to the mass concentration at the center of the region. In this
 case, gravity can potentially drive the accretion from the
 outskirts towards the center. This behavior is similar to what is found in
 Section \ref{sec:clump} for the simplified centrally condensed clump model.

 \subsection{IC348-B3 star-forming region}\label{sec:ic348}

 \begin{figure}
\includegraphics[width = 0.50 \textwidth]{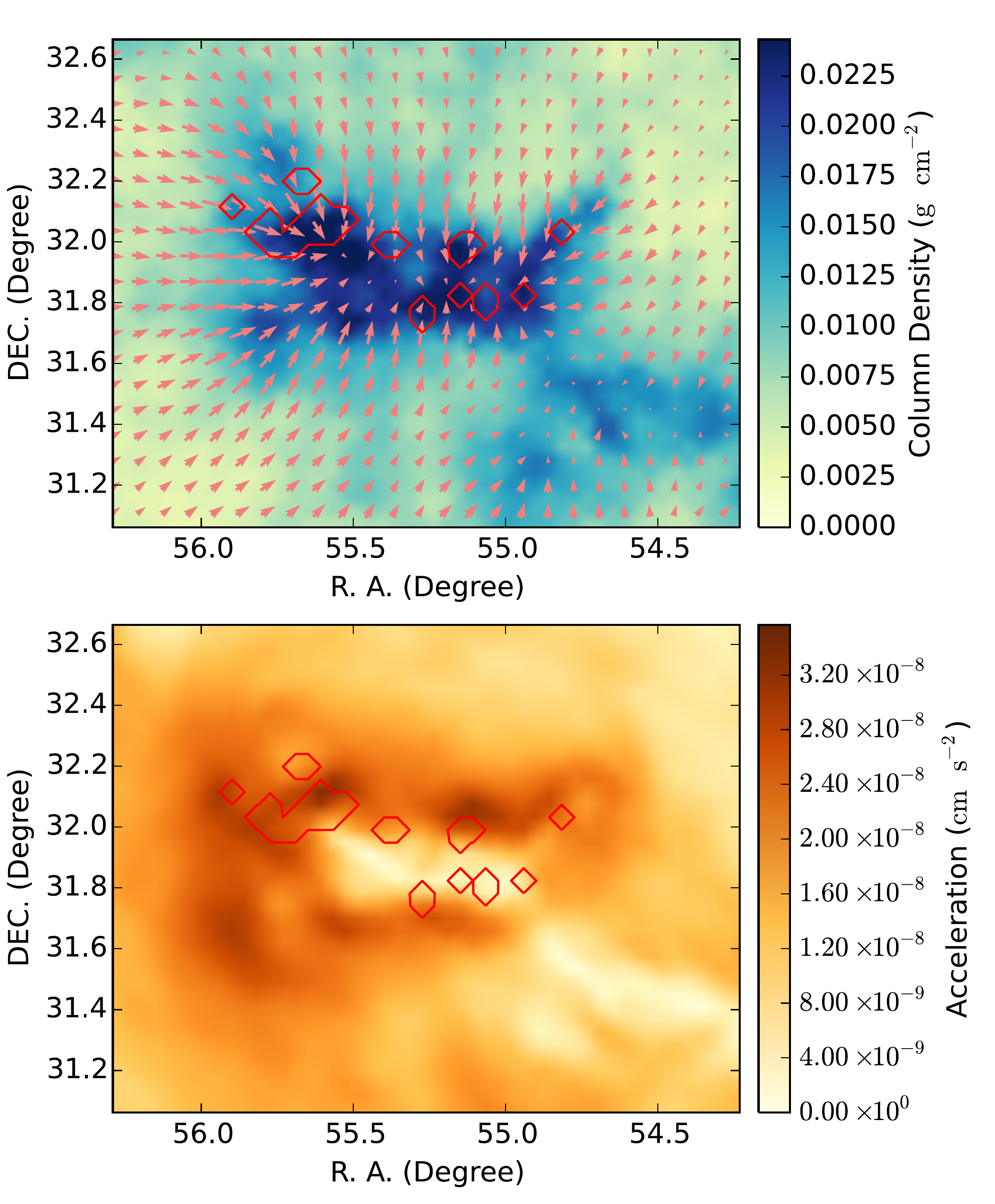}
\caption{{  Upper Panel:} surface density distribution and acceleration of the
IC348-B3 region in the Perseus molecular cloud. The background image is the surface density distribution and the
vectors stand for acceleration.
{  Lower Panel:} A map of the magnitude of acceleration of the same region.
\label{fig:ic348} The red contours mark the region where starless and
prestellar cores are found \citep{2006ApJ...638..293E}. Here one degree
correspond to $\sim 4.5\;\rm pc$. A color-blindness-proof
version of the this figure and be found in Appendix \ref{sec:appendix:a}.}
\end{figure}

 We consider acceleration in a highly irregular
 region -- the IC348-B3 star-forming region in the Perseus molecular cloud. The
 region is considered to be more evolved \citep{2009ApJS..184...18G}, and a
 young embedded cluster IC348 is found to be associated with the region.

In Fig. \ref{fig:ic348}  we present the results.
Positions of the dense cores are taken from \citet{2006ApJ...638..293E},
and they are marked with the red contours. Here acceleration concentrates at the
edges of the cloud. The region where acceleration is enhanced coincide with the
region where the dense cores are found. In this sense, the IC348
region resembles the  case of the ``Ghost'' studied in
\citet{2004ApJ...616..288B}. This resemblance points to a picture where the star
formation in the IC348 region traced by the dense cores are triggered because of the enhanced
gravitational acceleration at the edges of the region.

In the inner part of the IC348-B3 region, albeit the presence of gas,
the acceleration is reduced because of the apparent sheet-like geometry.
This lack of acceleration should lead to a lack of the ability of the gas to collapse and hence a lack of star
formation. Indeed, none of the dense cores in
\citet{2006ApJ...638..293E} are found at the inner part of the IC348-B3.
This is similar to the thin disk example as discussed in Section \ref{sec:disk}.

\section{Discussions}\label{sec:disc}
\subsection{General importance}\label{sec:compare}

 \begin{figure}[h]
\includegraphics[width =  0.50 \textwidth]{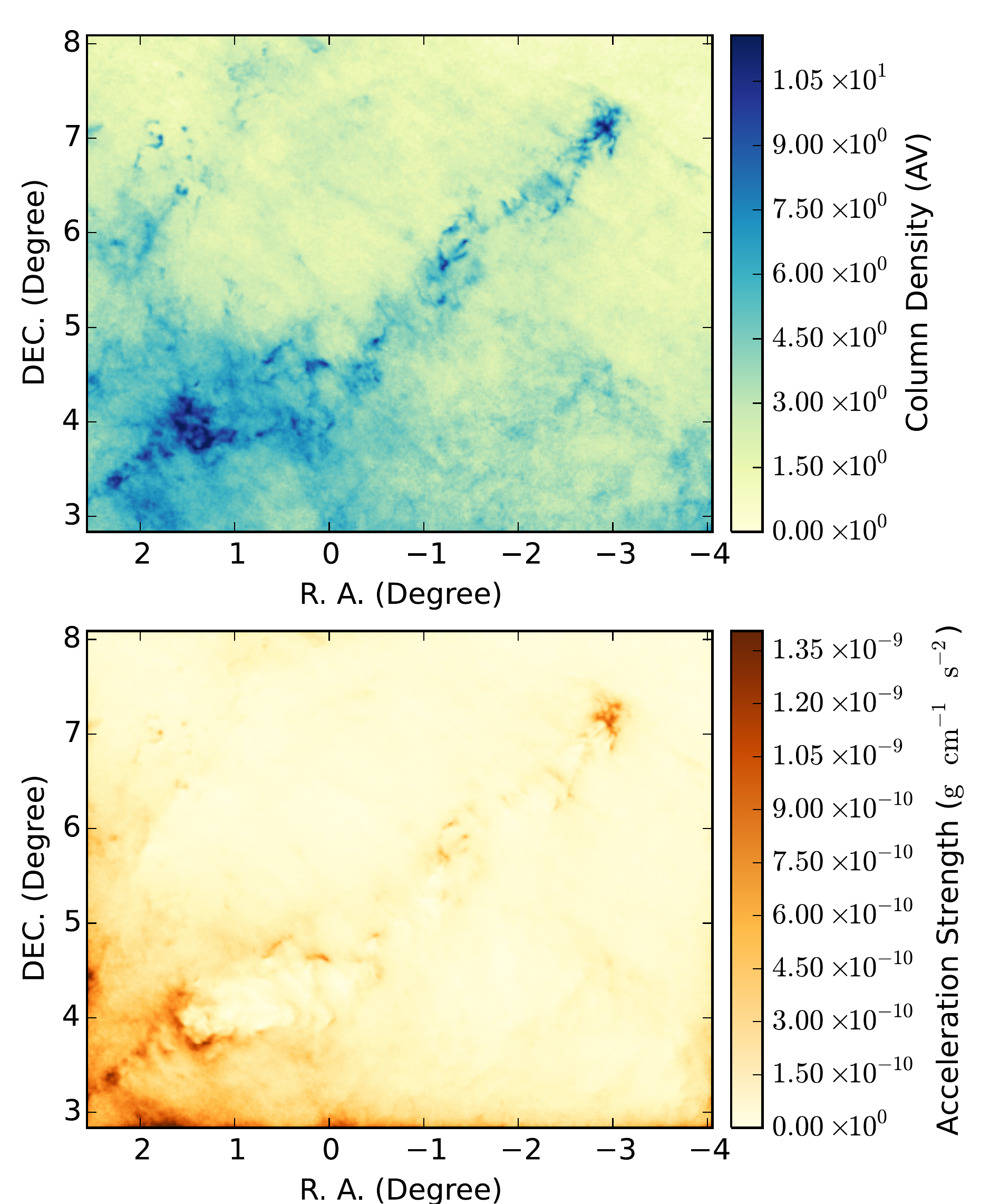}
\caption{{  Upper Panel:} A map of of surface density of the Pipe nebula.
{  Lower Panel:} A map of the magnitude of effective acceleration pressure
$p_{\rm acceleration} \sim \Sigma \; |\vec{a}|$. Here one degree correspond to $\sim
3.6 \;\rm pc$.
\label{fig:pipe:compare} }
\end{figure}

 \begin{figure}[h]
\includegraphics[width = 0.50 \textwidth]{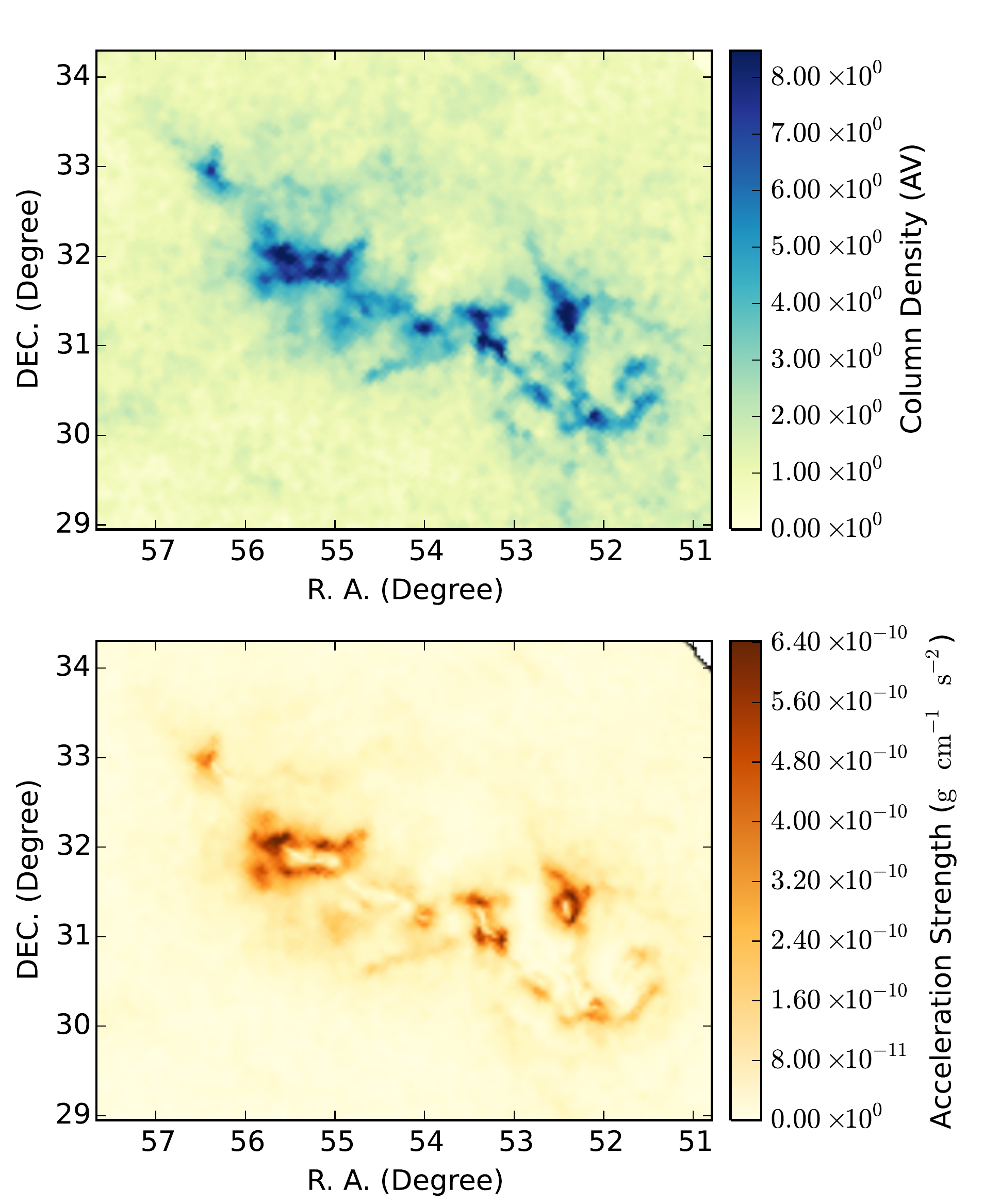}
\caption{{  Upper Panel:} A map of of surface density of the Perseus molecular
cloud.
{  Lower Panel:} A map of the magnitude of effective acceleration pressure
$p_{\rm acceleration} \sim \Sigma \; |\vec{a}|$. Here one degree correspond to $\sim
4.5 \;\rm pc$.
\label{fig:perseus:compare} }
\end{figure}
The acceleration map  can be combined with the map of surface density to
estimate the importance of gravity in accelerating the gas. This can be achieved
by multiplying $| \vec a|$ with the surface density map $\Sigma$. {  The
effective acceleration pressure is}
\begin{equation}\label{eq:p:accel}
 p_{\rm acceleration} = \frac{F_{\rm acceleration}}{S} = \frac{M \times
|\vec a_{\rm 3D}|}{S} \sim \Sigma \; |\vec a_{\rm 2D}|\;,
\end{equation}
where $\vec a_{\rm 3D}$ is the 3D acceleration and $\vec a_{\rm 2D}$ is the 2D
projected acceleration. We assume that $|\vec a_{\rm 3D}| \sim |\vec a_{\rm
2D}|$.

Eq. \ref{eq:p:accel} provides an order-of-magnitude estimate of
the importance of acceleration in molecular clouds, and can be
used to obtain maps of the importance of acceleration on
the sky plane.

In the localized regions, {  the effective acceleration pressure }can
reach $ p \sim 10^{-9} \rm g\; cm^{-1}\; s^{-2}$, which correspond to
$\dot p/ k_{\rm B} \sim 10^{7}\; \rm K \; cm^{-3}$ where $k_{\rm B} $ is the
Boltzmann's constant.
This is much larger than the external ambient pressure of molecular clouds
estimated in previous works \citep[][]{2009ApJ...699.1092H,2011MNRAS.416..710F} \footnote{In
\citet{2011MNRAS.416..710F}, the cloud ambient pressure is estimated to range
from $10^{4}$ to $10^{7}$ $\rm K\;cm^{-3}$. Our estimate stays on the upper
end. For the Pipe Nebula, \citet{2008ApJ...672..410L} estimated a pressure of
$7\times 10^4 \rm K\;cm^{-3}$ for the whole region. This pressure is small (by
about two orders of magnitudes) compared to the  effective gravitational
acceleration pressure.}.

{
The comparison between the effective pressure and the ambient pressure provides
estimates on the importance of gravitational acceleration. However, because our
calculations are done in 2D rather than in 3D, the results needs to be
interpreted with caution.
The pressure is isotropic, and the momentum injection from acceleration is anisotropic. Consider a 3D cuboid of a size $(l_x, l_y, l_z)$ where $z$ stand for the direction of the line of sight,
and the acceleration points toward the $y$ direction. The momentum injection
from gravitational acceleration is $\dot P_{\rm accel} \approx \rho a_y  l_x l_y
l_z \approx \Sigma a_y \times l_x l_y \approx p_{\rm accel} l_x l_y $ where
$\Sigma$ is the surface density.
The momentum injection from the ambient pressure along the $y$ direction is
$\dot P_{\rm pressure} \approx p_{\rm ambient}l_x l_z $ where $p_{\rm ambient}$ is the ambient
pressure.
$\dot P_{\rm accel} / \dot P_{\rm pressure} \approx p_{\rm accel} / p_{\rm
ambient}$ requires $l_y \approx l_z$. In other words, the ratio between  $p_{\rm accel}$
and $p_{\rm ambient}$ is a meaningful representation of the relative importance
between acceleration and ambient pressure only when the region is only
moderately extended along the line of sight.
When the region is highly elongated along the line of sight, the importance
of the effective acceleration pressure will be over-estimated.

In our acceleration map, $p_{\rm accel}$ can be $ 100$ times stronger
than the local ambient pressure.
External pressure can be important of the cloud as a whole, but gravitational acceleration will be
extremely effective in the localized regions.

}

We also derive the timescale on which gravitational acceleration can change the
concentration of the gas significantly. Observationally, it is found that dense
gas concentrate in clumps and filaments, and the clumps are typically $\sim 1\rm
pc$ large \citep{2000A&A...355..617M,2002ApJ...566..945B}. For acceleration
to create such gas clumps, we have ({  using $l = 1/2\; a\; t^2$ where $l$ is
the size of the clump, $a$ is the acceleration and $t$ is the time})
\begin{equation}
t_{\rm clump} \approx \sqrt{\frac{l}{a}}\;,
\end{equation}
{  For the Pipe nebula, $a \sim 10^{-8}\;\rm cm\, s^{-2}$, from which we
estiamte a  timescale of $t_{\rm clump} \sim 10^6\;\rm yr$.}
$t_{\rm clump}$ is about one order of magnitude shorter than typical cloud
lifetimes, which are $\sim 10^7\;\rm yr$
\citep{1997ApJ...476..166W,2007prpl.conf...81B}, and is much shorter than the
global dynamical timescale of the Pipe nebula, which is around $t_{\rm dyn} \sim
\sqrt{r^3 / G m} \sim\, 3 \times 10^7 \rm yr$ {  where r is the size of the
cloud. We used $L = 15 \;\rm pc$ and $m=10^4 M_{\odot}$ }.
Neglecting other factors such as magnetic fields,
gravitational acceleration can potentially trigger fast collapses in the
localized regions.

\subsection{Diverse behaviors}

 \begin{figure*}[!htb]
\includegraphics[width = 0.9   \textwidth]{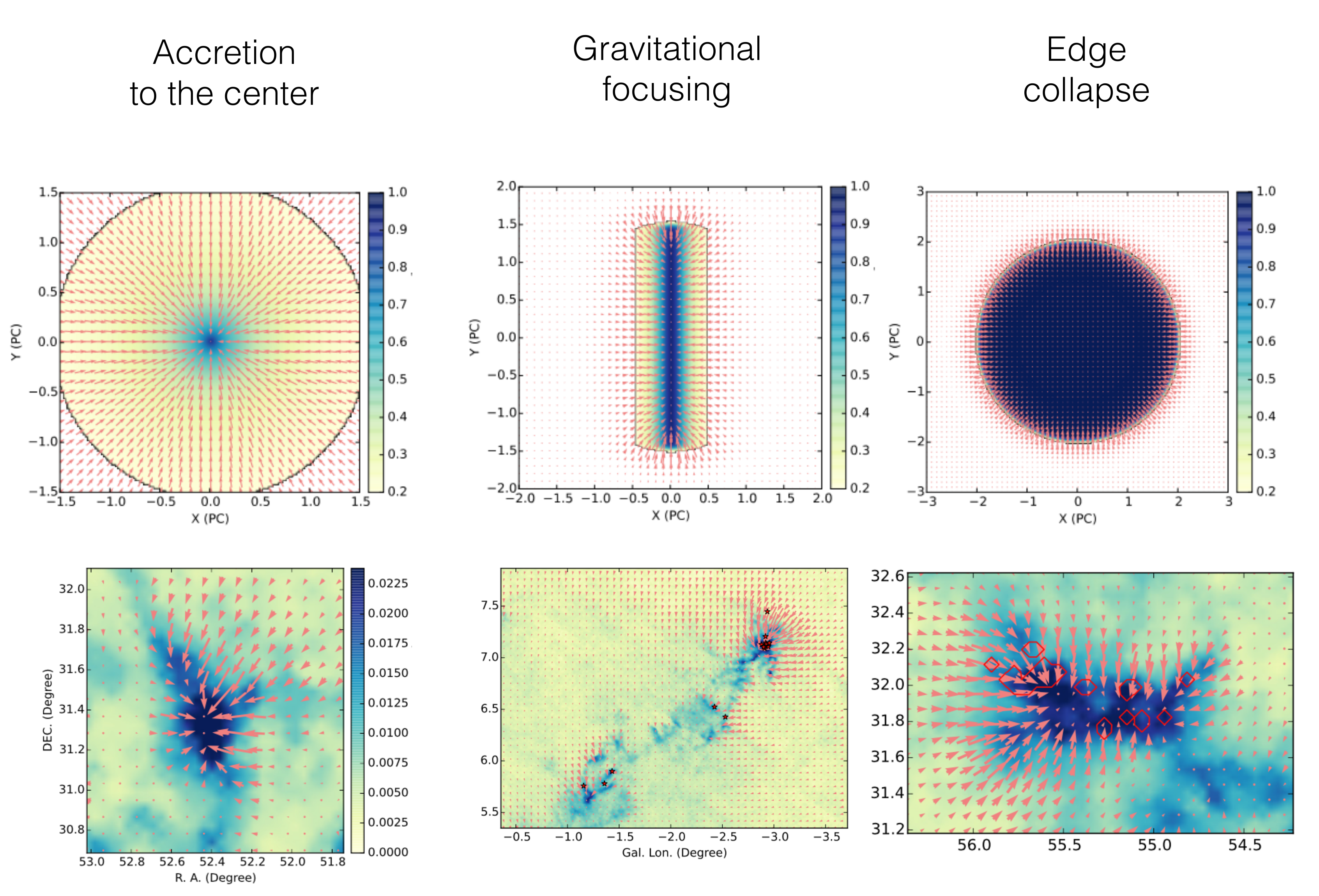}
\caption{\label{fig:summary} A summary of the effects discussed in this work.
The first row lists the names of the effects, the second row shows the idealized
models, and the third row shows the relevant observations. The first model on
the second row is discussed in Section \ref{sec:clump},
the second model on the second row is discussed in Section
\ref{sec:filament}, and the third model on
the second row is discussed in \ref{sec:disk}. The first observation
on the third row is the  NGC1333 region (Sec. \ref{sec:ngc1333}), the second
observation is the  Pipe nebula (Sec. \ref{sec:pipe}), and the third observation
is the IC348-B3 region (Sec. \ref{sec:ic348}).}
\end{figure*}

We find that gravitational acceleration exhibits diverse behaviors in various
regions, which can be summarized in Fig. \ref{fig:summary}.

In the Pipe nebula, we find that acceleration concentrates at the B59 region.
The acceleration structure of the Pipe nebula can be explained by the
``gravitational focusing'' effect proposed by \citet{2004ApJ...616..288B}.
Similar phenomena are seen in many of the
recently-observed filaments
\citep{2015Natur.520..518L,2015arXiv150703742K,2015arXiv151007063B,2014MNRAS.439.3275W},

 We also
 the method
to the Perseus molecular cloud, and find that in the IC348-B3 region both the
acceleration and the formation of dense gas tend to occur at the edges of
the region, and this is consistent with the ``edge collapse'' in
\citet{2004ApJ...616..288B}. Acceleration in the NGC1333 cluster-forming
region converges towards the center, indicating that gravity can collect
 matter towards the center of the region and feed star formation. In sum, we
 that
gravitational acceleration plays important roles at least in the following
ways:
\begin{enumerate}
 \item In driving the accretion onto the central parts of centrally-condense
regions (accretion).
\item In driving the collapse at the ends of the filamentary structures
(gravitational focusing).
\item In driving the collapse at edges of flat structures (edge collapse).
\end{enumerate}
The various ways in which gravitational acceleration functions are summarized in
Fig. \ref{fig:summary}. It should be noted that in many cases these different
mechanisms occur simultaneously. One example is the gas accretion at the ends
of a filament.
In this case, gravitational focusing is responsible for the collapse of the
filament ends, and if a diffuse envelope exists around the filament, accretion
will occur which brings gas from the diffuse phase to the  ends of the filament
\citep{2013ApJ...769..115H}.

\section{Conclusions \& Discussions}\label{sec:conclusec}
In this work, we present a gravitational acceleration mapping method to
quantify the importance of gravity in molecular clouds, and study the impact of
gravitational acceleration on the evolution of molecular clouds.

The method takes observational maps of
surface density distributions on the sky plane as the inputs, and provides maps
of the gravitational acceleration on the sky plane.
It provides a qualitative and intuitive picture of gravity in molecular
clouds. We apply
 our method to both simplified models and observations including
Pipe nebula and the Perseus molecular cloud.

With observational data, we find that the  star formation activities in
the IC-348 region traced by dense cores can be explained by edge collapse, and
the concentration of star formation in the B59 region of the Pipe nebula is
consistent with a picture where gravitational focusing contributes to the
compression of the gas.  In the NGC1333 cluster-forming region, acceleration can
drive accretion from the outskirts to the center.
The main observational results are summarized as follows:
\begin{itemize}
  \item Gravitational acceleration exhibits diverse behaviors in various
  regions.
  \item Gravitational acceleration tends to concentrate in the localized regions
  in molecular clouds.
  \item  In the localized regions, the strength of the acceleration exceeds by
  much the mean external pressure of molecular clouds.
  \item In the localized regions, the concentration of acceleration can lead to
  local collapse. The local collapse can occur in a timescale that is much
  shorter (e.g. $10^6$ yr) than the dynamical timescale of the cloud.
\end{itemize}

However, a complete picture of gas evolution is yet to be achieved since we need
to fully understand the interplay between magnetic field, turbulence and gravity
in these these regions. There are ample observational
evidences that magnetic fields are important in star-forming regions
\citep{2014prpl.conf..101L}. Remarkably-ordered magnetic fields have been revealed in several star-forming regions
\citep{2014ApJ...794L..18Q, 2014ApJ...792..116Z,2008A&A...486L..13A,
2011ApJ...741...21C,2015arXiv150204123P,2015Natur.520..518L}. Magnetic fields
are observed to be either parallel or perpendicular to filamentary structures
\citep{2014ApJ...792..116Z,2015arXiv150204123P}. Magnetic field strength can
also vary strongly within different star-forming regions
\citep[e.g.][]{2008A&A...486L..13A}.
Strong magnetic field can affect fragmentation significantly
\citep{2015MNRAS.452.2410S}. The effect of turbulence is not considered explicitly in this formalism. It is
possible that in regions such as B59, turbulent motion can be driven as a
result of the enhanced acceleration \citep{2010A&A...520A..17K}. These
possibilities should be investigated in subsequent works.

The Milky Way ISM is in general structured. We expect that the non-local effect
of gravity to be important in the majorities of the cases.  The acceleration
mapping method presented in this paper provides a detailed and intuitive picture
of gravity in the molecular ISM. The results can be combined with studies of
turbulent motions and magnetic field observations to finally achieve a clear
picture of star formation.

\begin{acknowledgements}
Guang-Xing Li acknowledge supports from the priority
program SPP 1573 ``ISM-SPP: Physics of the Interstellar
Medium''. The paper made use of the \texttt{astropy} \citep{2013A&A...558A..33A}
package. Guang-Xing Li thanks Xun Shi for discussions. We thank the referee
who helped us to improve the paper significantly, and inspired us to rethink
some of our questions. The referee is also acknowledged
for proposing the cuboid model which we adopt in Sec. \ref{sec:compare}.
\end{acknowledgements}

\bibliography{paper}

\begin{appendix}

\section{Colorblindness-proof version of some of the
figures}\label{sec:appendix:a} 
\begin{figure*}
\includegraphics[width = \textwidth]{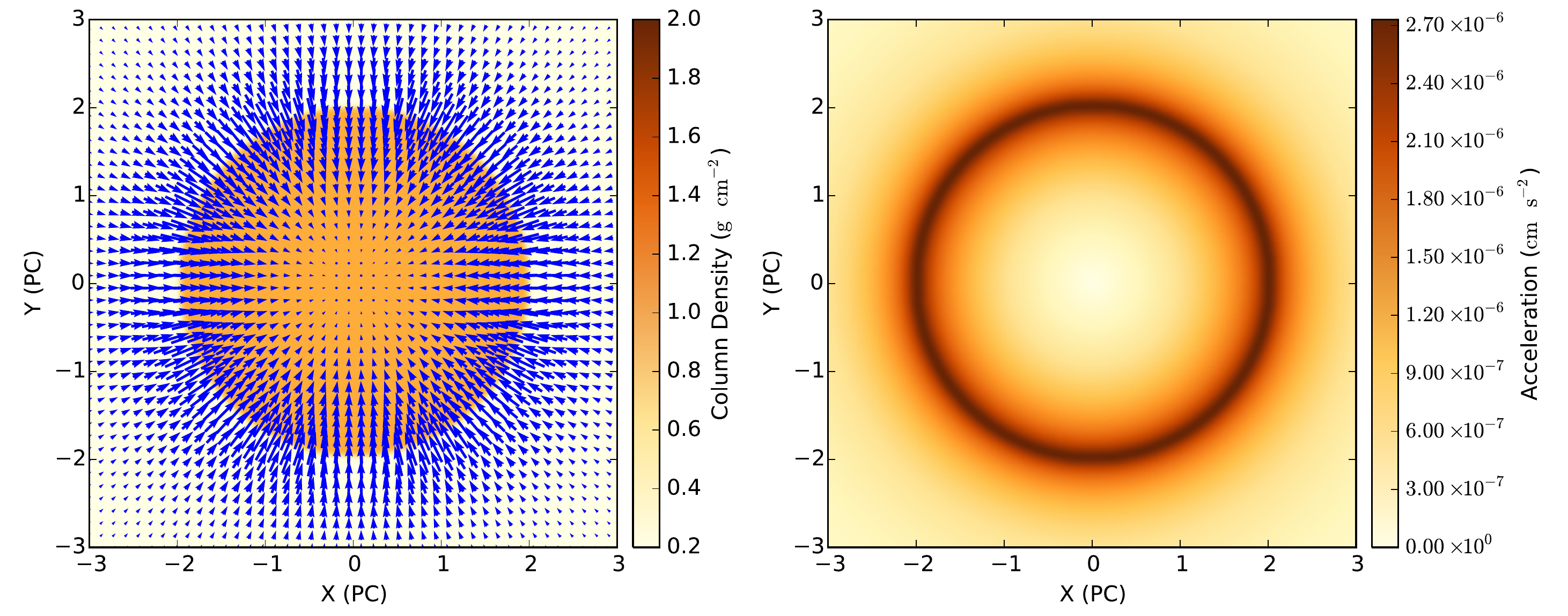}
\caption{{  Left Panel:} Density distribution and acceleration for our disk
model with a radius of $1\;\rm pc$ and a
surface density of $1\;\rm g \;\rm\; cm^{-2}$. The background image is the
density distribution and the vectors stand for acceleration. {  Right
Panel:} A map of the magnitude of acceleration.
\label{fig:disk:cb} }
\end{figure*}

In this section we provide plots that are
suitable for those with color-blindness. Fig. \ref{fig:disk:cb} correspond to Fig. \ref{fig:disk}.

\begin{figure*}
\includegraphics[width = \textwidth]{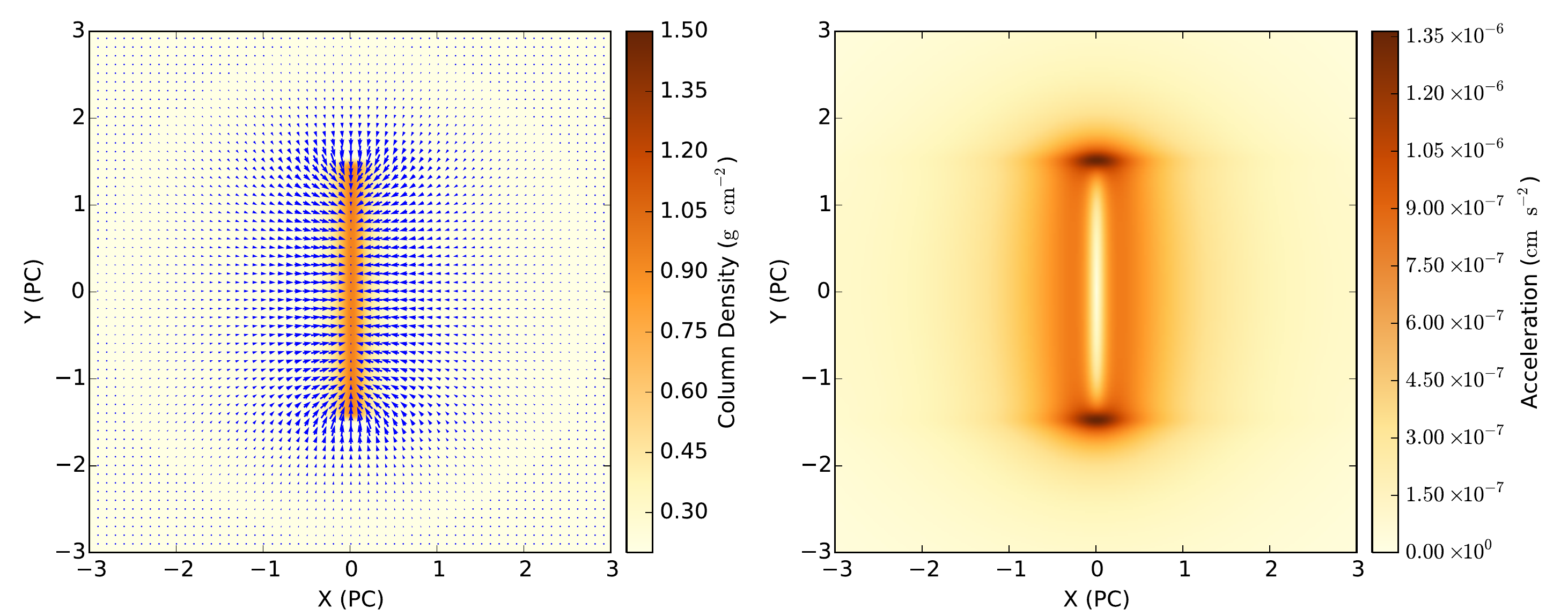}
\caption{{  Left Panel:} Density distribution and acceleration for our
truncated filament model (see Section \ref{sec:filament} for details). The
background image is the density distribution and the vectors stand for acceleration.
{  Right Panel:} A map of the magnitude of acceleration.
\label{fig:filament:cb} }
\end{figure*}

Fig. \ref{fig:filament:cb} correspond to Fig. \ref{fig:filament}.

\begin{figure*}
\includegraphics[width = \textwidth]{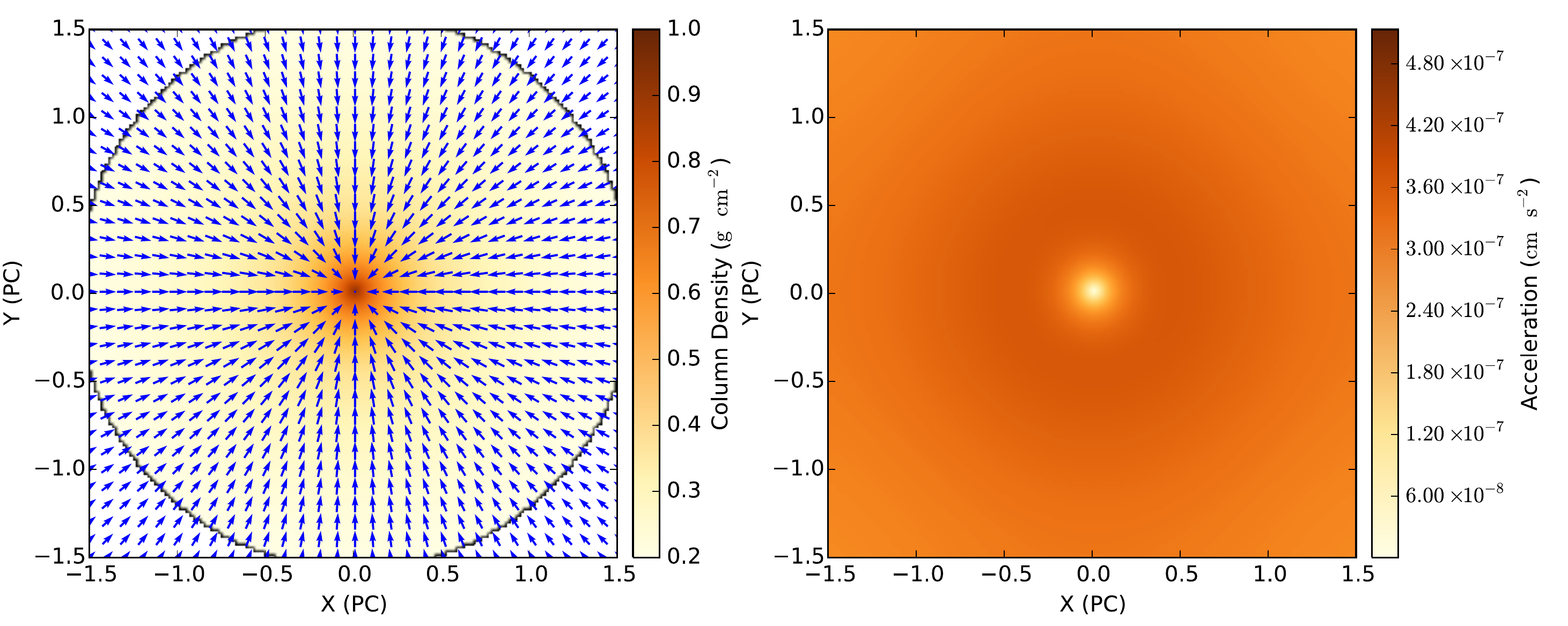}
\caption{{  Left Panel:} Density distribution and acceleration for our clump
model (see Section \ref{sec:clump} for details). The background image
is the density distribution and the vectors stand for acceleration. {  Right
Panel:} A map of the magnitude of acceleration.
\label{fig:clump:cb} }
\end{figure*}

Fig. \ref{fig:clump:cb} correspond to Fig. \ref{fig:clump}.

\begin{figure*}
\includegraphics[width =  0.9 \textwidth]{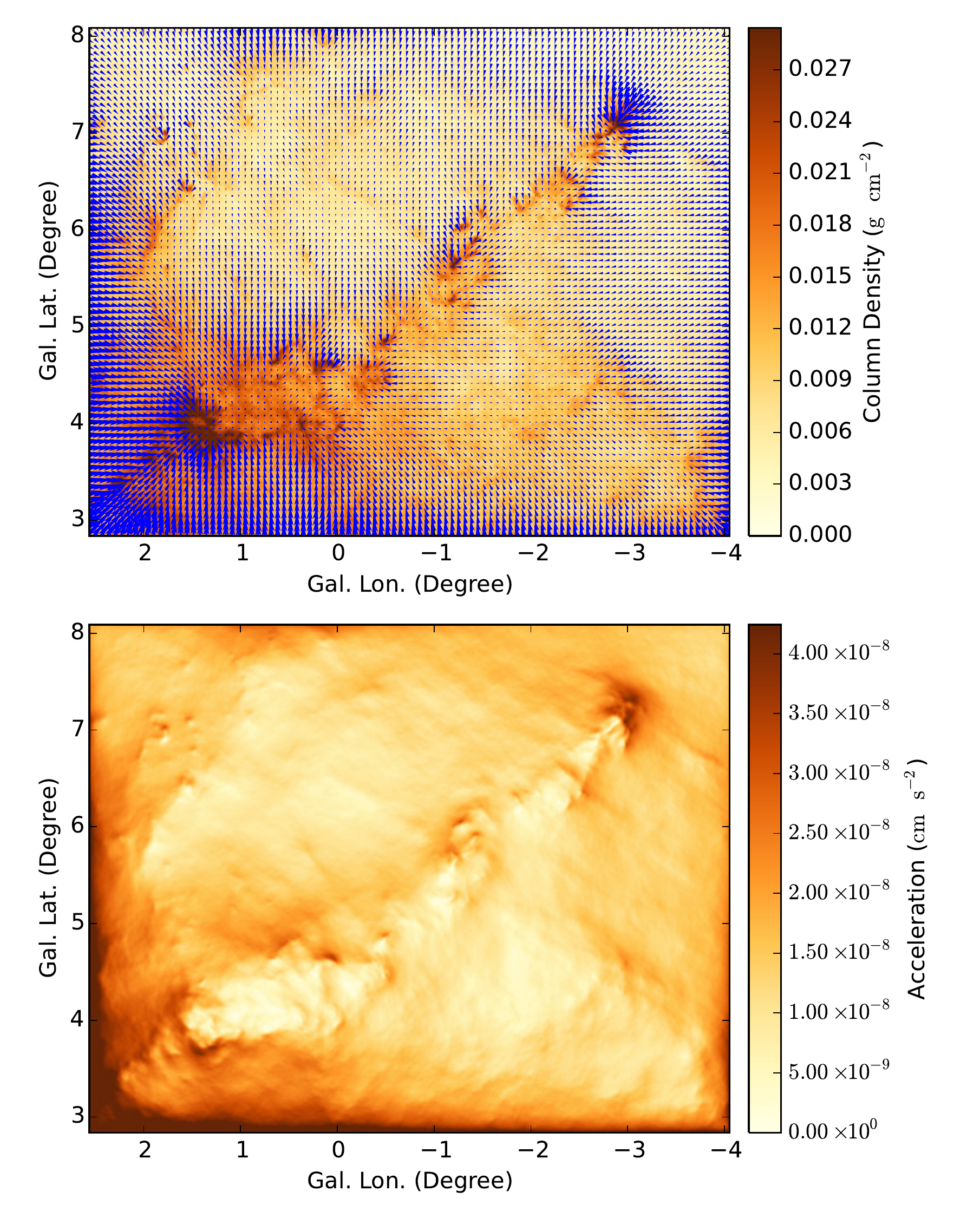}
\caption{{  Upper Panel:} surface density distribution and acceleration of the
Pipe region. The background image is the surface density distribution and the
vectors stand for acceleration. {  Lower Panel:} A map of the magnitude of
acceleration of the same region.  Here one degree correspond to $\sim 3\;\rm pc$.
\label{fig:pipe:region:cb} }
\end{figure*}

Fig. \ref{fig:pipe:region:cb} correspond to Fig. \ref{fig:pipe:region}.

\begin{figure*}
\includegraphics[width = 0.8 \textwidth]{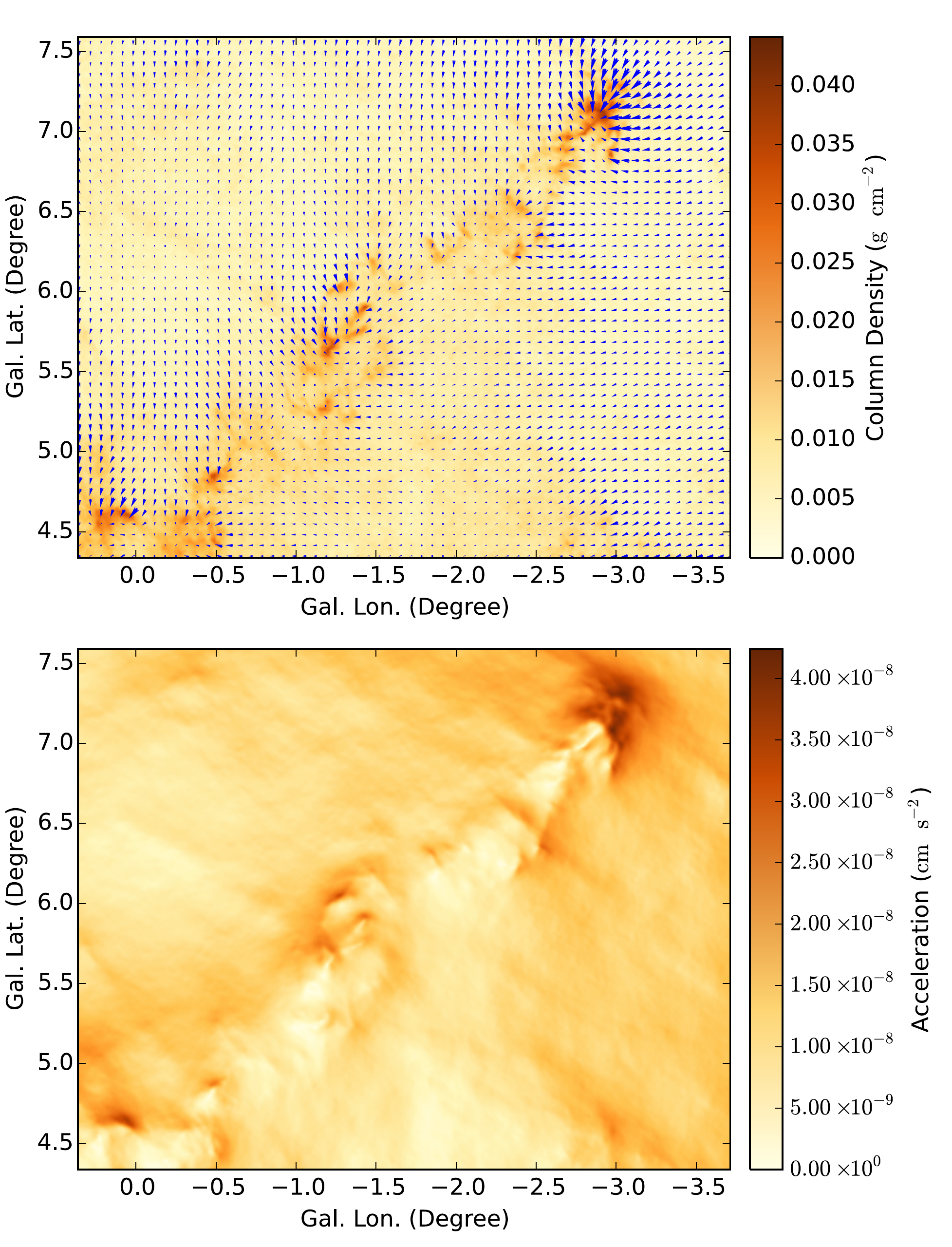}
\caption{{  Upper Panel:} surface density distribution and acceleration of the
Pipe Nebula. The background image is the surface density distribution and the
vectors stand for acceleration.
{  Lower Panel:} A map of the magnitude of acceleration of the same region.
The red stars stand for the YSO candidates in \citep{2009ApJ...704..292F}.  Here one degree correspond to $\sim
3\;\rm pc$.
\label{fig:pipe1:cb} }
\end{figure*}

Fig. \ref{fig:pipe1:cb} correspond to Fig. \ref{fig:pipe1}.

\begin{figure*}
\includegraphics[width = \textwidth]{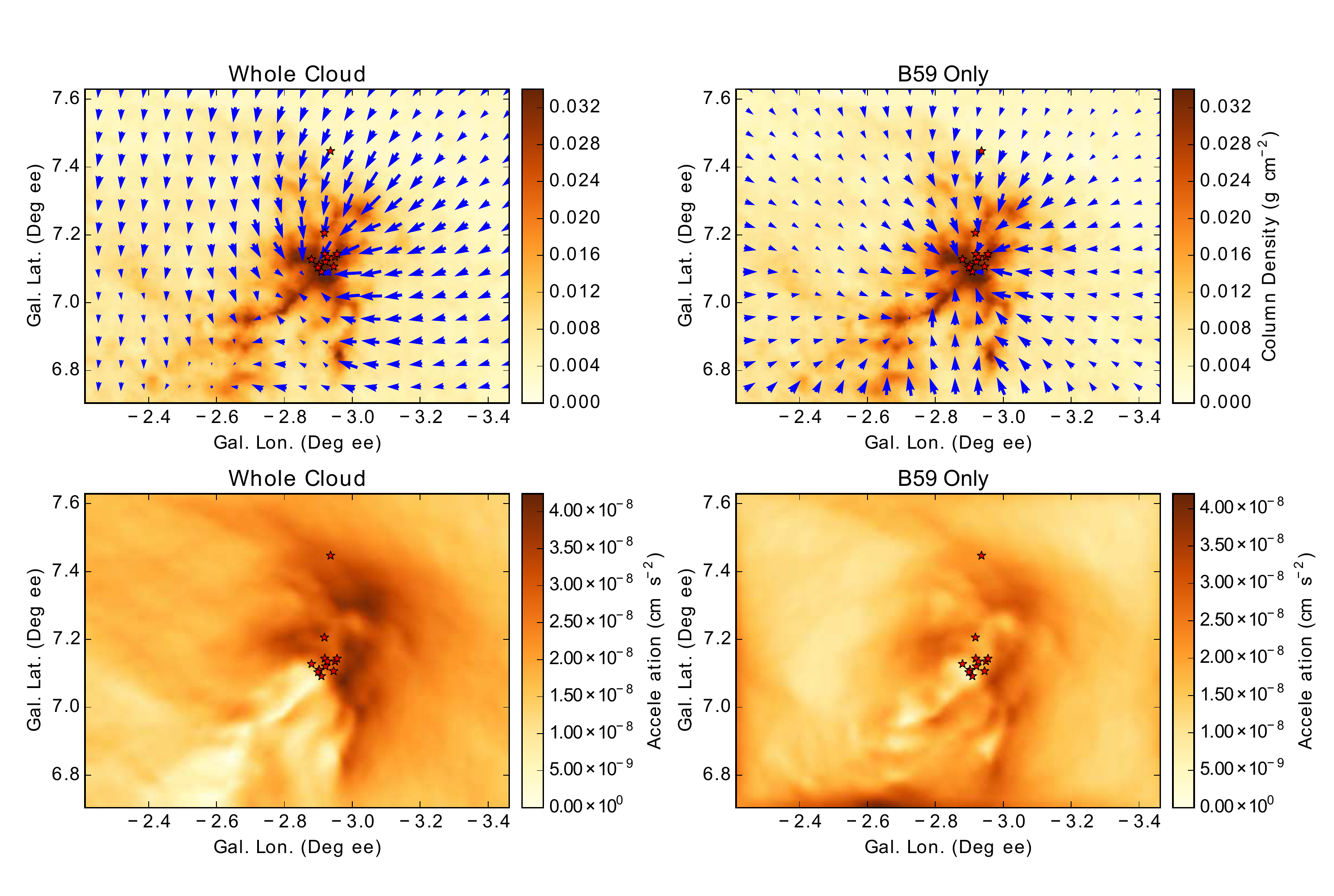}
\caption{{  Upper Panels:} surface density maps and acceleration of the
B59 region in the Pipe Nebula. The background images are the surface density
distributions and the vectors stand for acceleration.
{  Lower Panels:} Maps of the magnitude of acceleration of the same region.
The red stars stand for the YSO candidates in \citet{2009ApJ...704..292F}.
The results shown in the left panels are computed by taking all the
matter of the Pipe nebula into account \emph{(whole cloud)}, and the
results shown in the right panels are computed by taking only the matter
inside the B59 into account \emph{(B59 only)}.  Here one degree correspond to
$\sim 3\;\rm pc$. \label{fig:pipe2:cb} }
\end{figure*}

Fig. \ref{fig:pipe2:cb} correspond to Fig. \ref{fig:pipe1}.

\begin{figure*}
\includegraphics[width = 0.9 \textwidth]{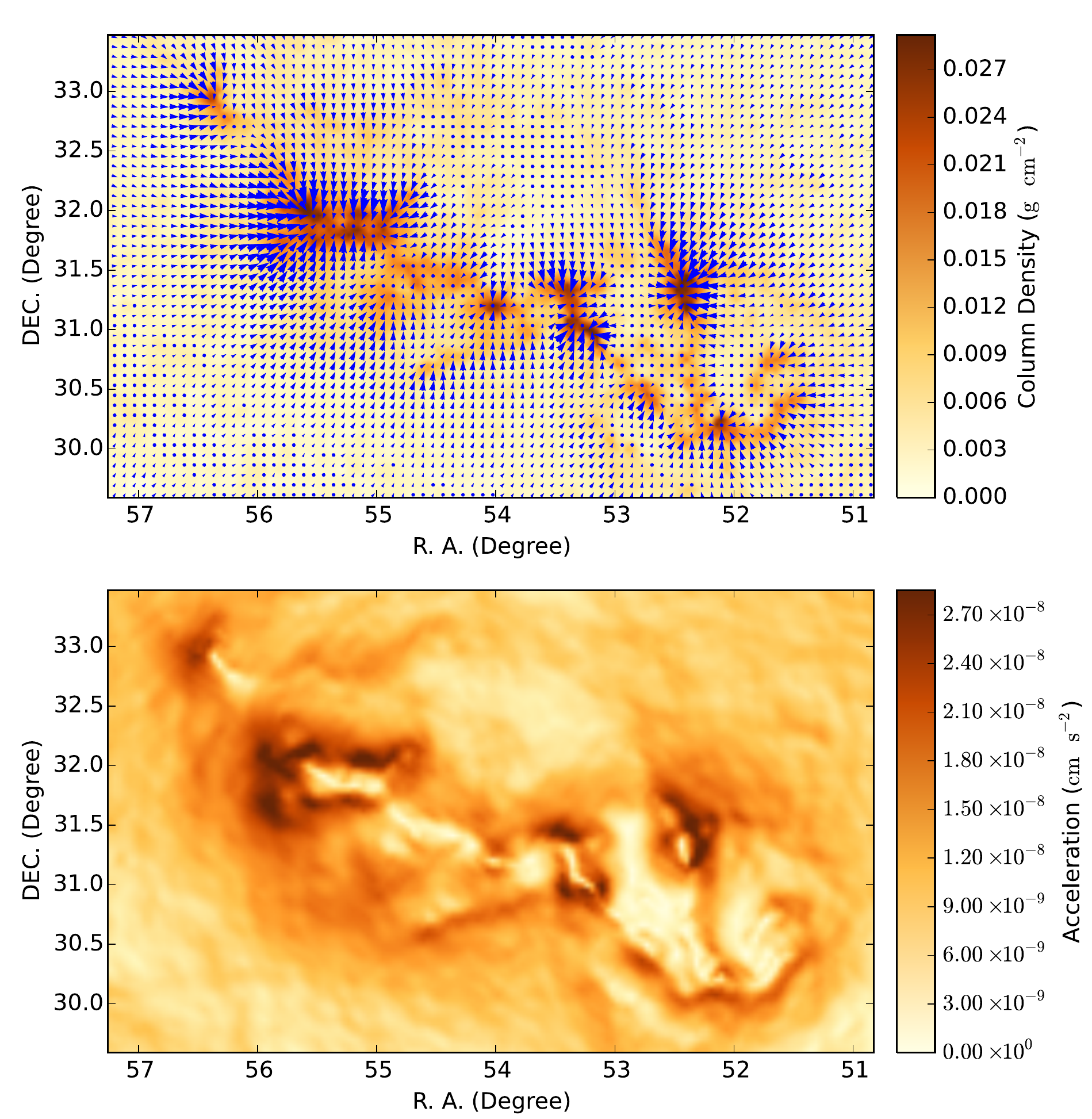}
\caption{{  Upper Panel:} surface density distribution and acceleration of the
Perseus molecular cloud. The background image is the surface density distribution and the
vectors stand for acceleration.
{  Lower Panel:} A map of the magnitude of acceleration of the same region.
Here one degree correspond to $\sim 4.5\;\rm pc$.
\label{fig:perseus_all:cb} }
\end{figure*}

 Fig. \ref{fig:perseus_all:cb} correspond to Fig. \ref{fig:perseus_all}.

\begin{figure*}
\includegraphics[width = \textwidth]{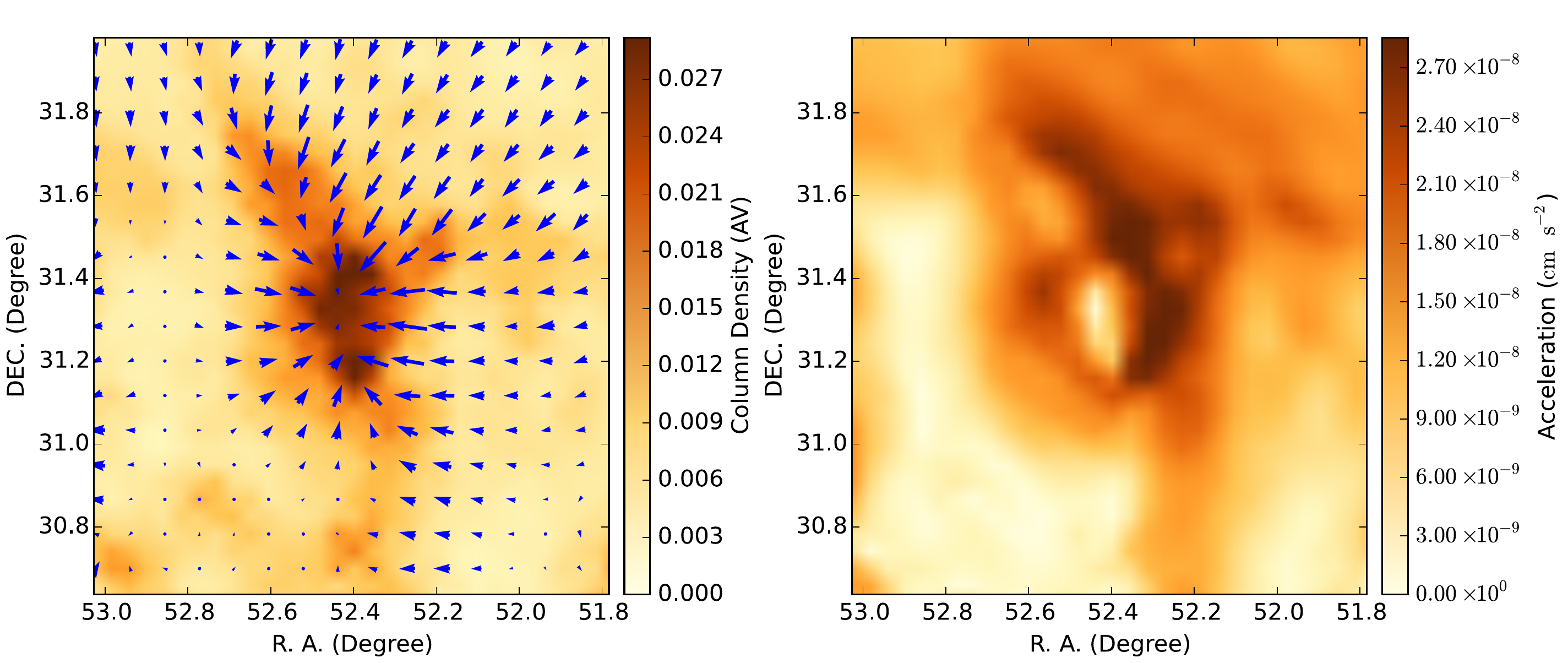}
\caption{{  Left Panel:} surface density distribution and acceleration  of the
NGC1333 region in the Perseus molecular cloud. The background image is the surface density distribution and the
vectors stand for acceleration.
{  Right Panel:} A map of the magnitude of acceleration of the same
region. Here one degree correspond to $\sim 4.5\;\rm pc$.
\label{fig:ngc1333:cb} }
\end{figure*}

 Fig. \ref{fig:ngc1333:cb} correspond to Fig. \ref{fig:ngc1333}.

\begin{figure}
\includegraphics[width = 0.48 \textwidth]{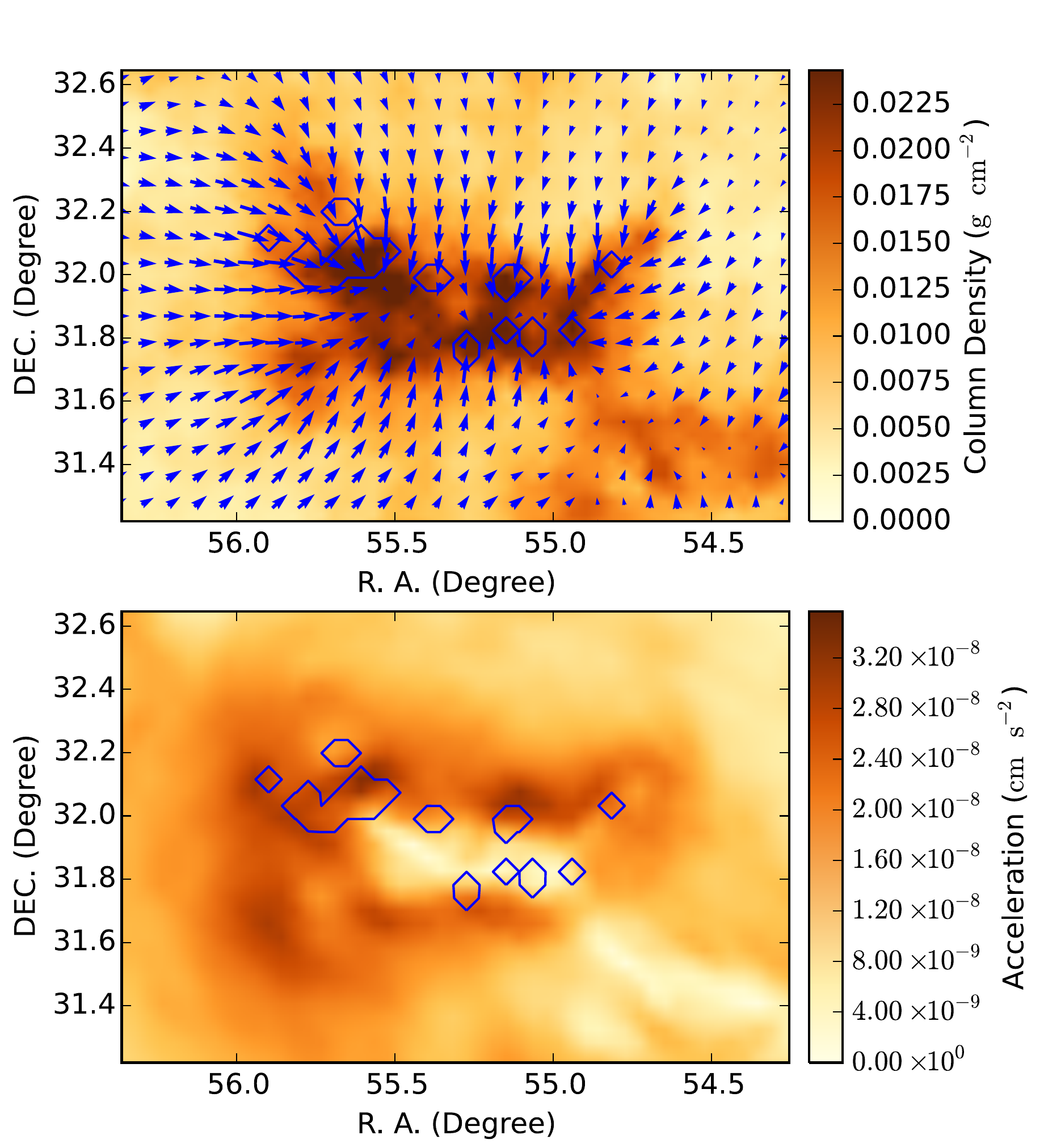}
\caption{{  Upper Panel:} surface density distribution and acceleration of the
IC348-B3 region in the Perseus molecular cloud. The background image is the surface density distribution and the
vectors stand for acceleration.
{  Lower Panel:} A map of the magnitude of acceleration of the same region.
\label{fig:ic348:cb} The blue contours mark the region where starless and
prestellar cores are found \citep{2006ApJ...638..293E}. Here one degree
correspond to $\sim 4.5\;\rm pc$.}
\end{figure}

  Fig. \ref{fig:ic348:cb} correspond to Fig. \ref{fig:ic348}.

\end{appendix}

\end{document}